\documentclass{article}
\usepackage{amsmath} 
\usepackage{amsfonts}
\usepackage{graphicx}
\usepackage{breqn} 
\usepackage{tikz}
\usetikzlibrary{decorations.markings}
\usetikzlibrary{calc}
\usepackage[all,cmtip]{xy}
\usepackage{tikz-cd}
\usepackage{enumitem}
\usepackage{cite}

\usepackage{amsthm}
\theoremstyle{definition}
\newtheorem{definition}{Definition}[section]

\usepackage[margin=1in]{geometry}
\usepackage{hyperref}

\title{\Huge  Spectral Geometry with Exceptional Symmetry and Charged Higgs Fields}
\author{S. Farnsworth$^{a,b}$ \\ 
 	\small $^a$Department of Mathematics, University of Regensburg,\\ 
 	\small$^b$Max Planck Institute for Gravitational Physics (Albert Einstein Institute), Germany.
 }

\begin{document}

\maketitle

\vspace{.5cm}

	\begin{abstract}
	We lay the foundations for a general approach to nonassociative spectral geometry as an extension of Connes' noncommutative geometry by explaining how to construct finite-dimensional, discrete spectral geometries with exceptional symmetry, and gauge covariant Dirac operators. We showcase an explicit construction of a geometry corresponding to the internal space of a $G_2\times G_2$ gauge theory with charged scalar content and scalar representations restricted
by novel conditions arising from the associative properties of the coordinate algebra. Our construction motivates a new definition of bimodules over nonassociative algebras and a novel form of bimodule over semi-simple octonion algebras.  
	\end{abstract}
 
\section{Introduction} 

Connes’ noncommutative spectral geometry~\cite{Connes:1994kx} is fundamentally built on the use of modules.   Each geometry is specified by a coordinate algebra $A$ faithfully represented on a Hilbert space $H$ (i.e. a module over $A$), along with a Dirac operator  $D$ that provides metric information. Bimodules  $\Omega_D^i A\in End(H)$ over the coordinate algebra $A$ are then constructed from this data, which function as a spaces of differential forms~\cite{Connes:reconstruction,Connes:2008kx}. Because modules are so central to the formalism, and because the representation theory of nonassociative algebras is significantly more challenging, fragmented, and less well developed than that of associative algebras, it is often assumed that spectral geometry must be limited to associative coordinate algebras. This assumption is, however,  incorrect. We have explicitly demonstrated that spectral geometries can be constructed for both special~\cite{Boyle:2020,Farnsworth_2020,Besnard_2022} and exceptional~\cite{farnsworth2025npointexceptionaluniverse} Jordan algebras, as well as for simple models based on the octonions~\cite{Farnsworth:2013nza}.

When constructing nonassociative geometries corresponding to gauge theories, it is often technically simpler to build models with uncharged Higgs fields. An important exception arises in the case of special Jordan algebras, where one can construct charged Higgs fields in a manner closely analogous to the associative case~\cite{Besnard_2022}.  As we will explain in Sec.~\ref{Sec_Split_Alternative_Bimodules}, it is usually  simpler to construct uncharged Higgs fields in the more general nonassociative setting because many of the challenges that arise from the nonassociativity of the coordinate algebra can be avoided by constructing Dirac operators  from the tensor product of elements from the associative nucleus of the coordinate algebra such as the identity element. Conceptually, this trick ensures that the interactions between the coordinate algebra $A$ and Dirac operator $D$ of a spectral triple remain to a large extent associative. While this approach greatly facilitates the construction of nonassociative spectral geometries,  the associative nucleus is typically invariant under the internal symmetries of an algebra. As a result, Dirac operators built in this way often yield physical models in which the Higgs fields they parameterise  are singlets under the internal gauge group. This  was first exhibited in\cite{Besnard_2022}, where we constructed internal spaces of gauge theories with $F_4\times F_4$ gauge symmetry and singlet Higgs fields.

In this paper we address the full challenge of constructing discrete, finite-dimensional, nonassociative spectral geometries that describe the internal spaces of gauge theories with charged Higgs fields.  We focus on internal spaces coordinatized by non-simple octonion algebras, motivated by a key  property sometimes referred to as an associative specialization~\cite{jacob1968}. Specifically, if the left and right actions of the octonion algebra $\mathbb{O}$ on itself are defined by $L_ab=ab$  and $R_ab=ba$ respectively, then the following identity holds for all $a,b\in \mathbb{O}$:
\begin{align}
	L_{ab} = L_aL_b + [L_a,R_b].\label{specialization}
\end{align}
This rule is easily derivable for any alternative algebra, whereby the associator $[a,b,c]=(ab)c-a(bc)$ changes sign under the exchange of any two elements $a,b,c\in A$.  
The specialization identity plays a role analogous to the homomorphism property $\pi_{ab}=\pi_a\pi_b$ of  representations of associative algebras $\pi:A\rightarrow End(M)$,  and to the specialization rule $\pi_{ab} = \frac{1}{2}\{\pi_a,\pi_b\}$ for associative representations of special Jordan algebras~\cite{jacob1968}. These identities are used heavily when defining Dirac operators and differential forms in both associative and special Jordan spectral geometry~\cite{Besnard_2022}. As we will show in Section \ref{sec_example_geometry} the specialization identity given in Eq.~\eqref{specialization} facilitates the construction of differential forms in the alternative setting as well, and gives rise to a novel notion of bimodules over the octonions, which  allows for the construction of Dirac operators which transform covariantly with respect to the symmetries of a spectral geometry.

  This paper is organized as follows: In Section~\ref{Sec_Bimodule_Problem} we review the standard approach to defining bimodules over nonassociative algebras. We explain the cases relevant to spectral geometry in which the standard bimodule definition fails, and suggest a new definition that caters specifically to the construction of nonassociative spectral geometries. In Section~\ref{Sec_Split_Alternative_Bimodules} we review the construction of differential forms and gauge covariant Dirac operators in associative spectral geometry. We explain the use of bimodule homomorphisms in such constructions, and show how the limited form that  bimodule homomorphisms take in the alternative setting  gives rise in the most naive approach to Dirac operators that remain invariant under the automorphisms of discrete, finite dimensional, alternative geometries. In Section~\ref{sec_example_geometry} we construct an explicit example of an alternative geometry, which corresponds to the internal space of a $G_2\times G_2$ gauge theory. We begin this construction with the naive definition of 1-forms constructed as what we term split alternative bimodules, and show how this results in a model in which the Dirac operator remains uncharged under the $G_2\times G_2$ symmetry. We then make use of the  specialization rule given in Eq.~\eqref{specialization} to derive a new form of bimodule over alternative algebras, which gives rise to a geometry with a covariant  Dirac operator, corresponding to the internal space of a gauge theory with charged Higgs fields.  In Section \ref{Sec_bimodules} we formalise the findings of Section \ref{sec_example_geometry}, defining 
  reconstituted (or charged) derivation bimodules over semisimple octonion algebras, together with derivation compatible maps from which one is able to construct Dirac operators. Finally, in Section \ref{Sec_discussion} we
   collate the key lessons learned   
   through the examples of associative, special Jordan, exceptional Jordan, and alternative geometries,  and  discuss the extension to a completely general  nonassociative spectral geometry. In the Appendix we collect results that are well known, but useful for understanding this paper (such as the properties of the octonions), and as well as perform calculations which would otherwise disrupt the flow of the paper.

\section{Bimodules In Nonassociative Spectral Geometry}
\label{Sec_Bimodule_Problem}

\subsection{The Standard Definition Of Nonassociative Bimodules}
\label{SSec_Bimodule_Standard_Def}

There is a standard approach to defining bimodules in terms of square zero extensions that goes back to Eilenberg, MacLane, and Hoschild~\cite{eilenberg1945natural,hochschild1945cohomology,eilenberg1948}, with later work by  Beck~\cite{Beck}, that generalizes naturally to nonassociative algebras.

\begin{definition}\label{def:sze}
Let $A$ be a (possibly non-associative) algebra over a field $\mathbb{F}$. A Square zero extension~\cite{hochschild1945cohomology} of $A$ is a new algebra with vector space $B=A\oplus M$, and with multiplication given by:
\begin{align}
	(a,h)(a',h') = (aa',a\cdot h' + h\cdot a'),
\end{align}
for all $a,a'\in A$, $h,h'\in M$, where:
\begin{itemize}
	\item  the product $A\otimes A\rightarrow A$, as denoted by $aa'$ above is given by the original product on $A$,
	\item $\cdot:A\otimes M\rightarrow M$ and $\cdot:M\otimes A\rightarrow M$ define the left and right  $\mathbb{F}$-linear actions of $A$ on $M$, respectively,
	\item the map $M\otimes M\rightarrow B$ is the zero map.
\end{itemize}
\end{definition}

\begin{definition}\label{def:bimodule}
Let $C$ be a class of (possibly nonassociative) algebras over a field $\mathbb{F}$. If $A$ is in $C$, and $M$ is a vector space over $\mathbb{F}$, then the square zero extension $B=A\oplus M$ is called a bimodule for $A$ in $C$ if $B$ is in $C$~\cite{schafer1955}. For example, if  $A$ is associative and $B$ is  associative, then $M$ is called an associative bimodule over $A$. If $A$ is a Jordan, alternative, or Lie algebra, and if $B$ satisfies the corresponding Jordan, alternative, or Lie identities, then $M$ is referred to as a Jordan bimodule, alternative bimodule, or Lie bimodule respectively.	That is, a bimodule of a certain class is defined to inherit its  algebraic properties from the algebra over which it is defined. 
\end{definition}

~

Defining bimodules via square-zero extensions with inherited algebraic properties offers two clear benefits. First, for associative algebras this approach reproduces the standard notion of associative bimodules~\cite{Schafer}. Specifically, the left and right bimodule products establish left and right representations $\pi_L:A\rightarrow End(M)$ and $\pi_R:A\rightarrow M$ such that $\pi_L(a)h = a\cdot h$ and $\pi_R(a)h = h\cdot a$ for all $a\in A$, $h\in M$. When $B=A\oplus M$ is associative, these representations   satisfy:
\begin{align}
	\pi_L(ab) -\pi_L(a)\pi_L(b)&=0,\label{Eq_Assoc_Module_1}\\
		\pi_R(ab) -\pi_R(b)\pi_R(a)&= 0,\label{Eq_Assoc_Module_2}\\
			[\pi_L(a),\pi_R(b)] &= 0,\label{Eq_Assoc_Module_3}
\end{align}
$a,b\in A$, which are the standard defining identities of associative bimodules.

Secondly, defining bimodules in terms of square zero extensions with inherited algebraic properties ensures that the continuous internal symmetries of the algebra   $A$ extend naturally to the bimodule $M$.
In particular, the continuous inner automorphisms  $Aut(A)$ of $A$ are generated infinitesimally by its inner derivations $Der(A)$, the form of which depends on the algebraic properties of $A$. For example, the inner derivations of associative, Jordan, Lie, and alternative algebras take the following forms~\cite{Schafer}:
\begin{align}
	\delta_a &= L_a-R_a, & &(Associative),\label{deralt2assoc}\\
	\delta_{ab} &= [L_a,L_b], & &(Jordan),\label{Eq_Der_Jordan}\\
	\delta_a &= L_a, & &(Lie),\label{Eq_Der_Lie}\\
	\delta_{ab} &= [L_a,L_b]+[L_a,R_b]+[R_a,R_b], & &(Alternative),\label{Eq_Der_Alterative}
\end{align}
for $a,b\in A$. By defining $M$ via a square-zero extension $B=A\oplus M$ that inherits the associative properties of $A$,   the inner derivations on $B$ are guaranteed to retain the same  form. This, in turn, ensures that the inner derivations of $A$ continue to generate automorphisms of  $B$, providing a consistent   `lift' or extension of the inner symmetries of $A$ to $M$~\cite{Farnsworth_2015}.

In the context of spectral geometry, the primary drawback of defining bimodules as square-zero extensions with inherited associative properties is that the inner derivations of an algebra $A$ act internally on $A$ by definition. That is, once lifted to the extended algebra $B=A\oplus M$, the inner derivations $\delta\in Der(A)$  continue to act as $\delta:A\rightarrow A$,  and $\delta:M\rightarrow M$. An important construction in any spectral geometry is a bimodule of one-forms $M=\Omega_D^1 A$ defined over the coordinate algebra $A$, and a  key component  of this construction is an `external'  derivation $\Delta: A\rightarrow M$, defined in terms of a Dirac operator~\cite{landi1997}.
Throughout this paper we will use capital deltas `$\Delta$' to refer specifically to `external' derivations $\Delta:A\rightarrow M$, while we refer to any other derivation outside of this context by a lower case delta `$\delta$'.  While the conventional definition of a bimodule does not preclude the existence of  exterior derivations (as seen in the associative setting), it also does not guarantee them. In the nonassociative setting the usual bimodule definition  is  especially limiting: very often the only available modules defined via square-zero extensions with inherited associative properties are free modules of the form $M = A\otimes V$, for some vector space $V$, and equipped with the left and right actions~\cite{jacob1968,okubo,Zhevlakov}:
\begin{align}
	a \cdot (b \otimes v) &= ab \otimes v, \\
	(b \otimes v) \cdot a &= ba \otimes v.
\end{align}
$a\in A$, $b\otimes v\in M$. In such cases, any external derivation $\Delta:A\rightarrow M$ must inherit its structure directly from the derivations on $A$.  This situation arises, for instance, with the exceptional Jordan algebra, where only free bimodules are available~\cite{jacob1968}. In that context, defining one-forms as bimodules in the sense of Def.~\ref{def:bimodule} limits one to differential calculi   based entirely on inner derivations~\cite{carotenuto2019,DV:2016}. This is problematic in the context of finite dimensional, discrete geometries, where the role of the Dirac operator and the 1-forms it generates is to connect the points of the space. This is a  task that cannot be fulfilled by inner derivations, which act internal to the discrete points of a geometry~\cite{farnsworth2025npointexceptionaluniverse}.

\subsection{The Challenge of Constructing Differential Forms As Bimodules In Noncassociative Spectral Geometry}
\label{SSec_Forms_As_Bimodules}

In this section we take a closer look at the challenges involved in constructing nonassociative spectral geometries, defined in terms of spectral triples $T=(A,H,D)$, where $A$ is a coordinate algebra encoding topological information, $D$ is a Dirac operator containing metric data, and $H$ is a Hilbert space on which $A$ and $D$ are represented~\cite{Connes:1994kx}. 
The  construction of a spectral geometry involves at least three different kinds of modules:
\begin{enumerate}
	\item The Hilbert space $H$ of a spectral triple, which acts as a module (and sometimes a bimodule~\cite{ConnesJ}) over the coordinate algebra $A$ of the spectral triple, as well as over what is known as the universal calculus  $\Omega_d A = \oplus_i \Omega_d^i A$ constructed from $A$.
	\item Each space of universal $i$-forms $\Omega_d^i A$ is constructed as  a bimodule over $A$~\cite{landi1997}.
	\item The representation $\Omega_D^i A$ of each space $\Omega_d^i A$ on $H$ acts as bimodules over $A$.
\end{enumerate}
Furthermore, a bimodule  homomorphism $\pi:\Omega_d A\rightarrow \Omega_D A\in End(H)$, ties all three kinds of module together. The challenge of extending noncommutative spectral geometry to the nonassociative setting rests primarily in the consistent generalization  of each of the above modules and their relationships to one another. Consider the usual associative construction of  `universal' 1-forms $\Omega_d^1 A$. The space of  universal 1-forms defined over an \emph{associative}, unital algebra $A$, denoted $\Omega_d^1A$, is the vector space generated as a left (or equivalently right) $A$-module by the symbols  $d[a]$, for $a\in A$, and where $d$ satisfies:
\begin{align}
d[ab]= a\cdot d[b] + d[a]\cdot b \label{Eq_Leib_exterior}
\end{align}
$a,b\in A$, with  left and right bimodule actions $\cdot:A\otimes \Omega_d^1 A\rightarrow \Omega_d^1 A$ and $\cdot:  \Omega_d^1 A\otimes A\rightarrow  \Omega_d^1 A$. Associativity, together with the Leibniz rule  given in \eqref{Eq_Leib_exterior} ensures that a general element of $\Omega_d^1 A$ can be expressed as~\cite{landi1997}:
\begin{align}
	\omega = \sum a\cdot d[b],
\end{align}
where the sum is over elements $a,b\in A$.  An explicit construction of $\Omega_d^1 A$ over  unital, associative algebras is built as follows~\cite{landi1997}: First, one starts with the with the  space $M=A\otimes A$, which forms a natural bimodule over  $A$, with the following left and right actions:
\begin{align}
	\pi_L(a)(b\otimes c)&= ab\otimes c\in M\label{Eq_split_left_action}\\
	\pi_R(a)(b\otimes c)&= b\otimes ca\in M\label{Eq_split_right_action}
\end{align}
$a\in A$, $b\otimes c\in M$. Next consider the submodule of $M$ given by
\begin{align}
	&ker(m:M\rightarrow A), & m(a\otimes b) =ab.\label{1forms}
\end{align}
We refer to this submodule as $\Omega^1_\Delta A$. Any element $\omega\in \Omega^1_\Delta A$ is of the form $\sum_i a_i\otimes b_i$ with $\sum_i a_ib_i= 0$. We are therefore free to write $\omega = \sum_i \pi_L(a_i)(\mathbb{I}\otimes b_i - b_i\otimes \mathbb{I})$, where $\mathbb{I}$ is the identity element on $A$. We see that $\Omega^1_\Delta A$ is generated as a left module by elements of the form $a\otimes \mathbb{I} - \mathbb{I}\otimes a$, while an analogous argument from the right shows that these elements also generate $\Omega^1_\Delta A$ as a right module. Furthermore, if one  introduces the map $\Delta_{\mathbb{I}\otimes\mathbb{I}}: A\rightarrow \Omega_\Delta^1 A$ given by\cite[Ch. 3]{Bourbaki}:
\begin{align}
	\Delta_{\mathbb{I}\otimes\mathbb{I}}[a] =a\otimes \mathbb{I} - \mathbb{I}\otimes a\label{deri}
\end{align}
for $a\in A$ and identity element  $\mathbb{I}\in A$, then one notices that this map is linear, and acts as a derivation:
\begin{align}
	\Delta_{\mathbb{I}\otimes\mathbb{I}}[ab] &= 	\Delta_{\mathbb{I}\otimes\mathbb{I}}[a]\cdot b+	a\cdot \Delta_{\mathbb{I}\otimes\mathbb{I}}[b], & &\forall a,b\in A. \label{leibder}
\end{align}
There is then an isomorphism of bimodules $\Omega_d^1 A\simeq  \Omega_\Delta^1 A$, with $a\cdot d[b]\leftrightarrow \pi_L(a)(b\otimes \mathbb{I} - \mathbb{I}\otimes b)$. By identifying $\Omega_\Delta^1 A$ with $\Omega_d^1 A$, the differential $d:A\rightarrow \Omega_d^1 A$ is given by~\cite{landi1997} $d[a]= \Delta_{\mathbb{I}\otimes\mathbb{I}}[a]$, $a\in A$.

The naive approach to nonassociative  spectral geometry would be replicate the above standard construction of `universal' 1-forms over an algebra $A$ by starting with a bimodule that has  vector space $M=A\otimes A$, with bimodule action inherited from the left and right products of the algebra itself as in Eqs.~\eqref{Eq_split_left_action} and \eqref{Eq_split_right_action}. Unfortunately, such bimodules almost always fail the standard bimodule definitions for nonassociative algebras as outlined in Def.~\ref{def:bimodule}. For example, Jordan and alternative bimodules in the standard sense of Def.~\ref{def:bimodule} satisfy $[a,h] = 0$ and $[a,b,h] = [h,a,b]=-[a,h,b]= [b,h,a]$ respectively for $a,b\in A$, $h\in M$. These identities explicitly relate the left and right actions of an algebra $A$ on a bimodule $M$, while in contrast the left and right actions defined in  Eqs.~\eqref{Eq_split_left_action} and \eqref{Eq_split_right_action} are independent by construction. The standard bimodule definition in terms of a  square zero extension $B=A\oplus M$ with inherited  algebraic properties therefore appears to rule out the standard construction of universal calculi, which is foundational in noncommutative spectral geometry. The challenge deepens when investigating the construction of representations of the universal 1-forms $\pi:\Omega_d^1 A\rightarrow \Omega_D^1 A\in End(H)$.

The space of \textit{associative} 1-forms $\Omega_d^1 A$ is known as universal because given any associative  $A$ bimodule $M$, with derivation $\Delta:A\rightarrow M$, there exists a unique bimodule homomorphism $\pi:\Omega_d^1 A \rightarrow M$ such that  $\Delta = \pi\circ d$~\cite{landi1997}. When constructing associative spectral geometries the map:
\begin{align}
	\Delta_D(a)=\pi_L(d[a]) = [D,\pi_L(a)]\label{Eq._homomorphismD}
\end{align}
$a\in A$, with $D$  the Dirac operator corresponding to the spectral triple $T = (A,H,D)$, 
defines a bimodule homomorphism $\pi_L:\Omega_d^1 A\rightarrow \Omega_D^1 A\in End(H)$, and   extends the algebra representation $\pi_L:A\rightarrow End(H)$ to include  1-forms  $\pi_L:A\oplus \Omega_d^1 A\rightarrow End(H)$. The space $\Omega_D^1 A$ acts naturally as a bimodule over $A$, with the left and right actions inherited from the operator product in $End(H)$ (modulo `junk' forms~\cite{landi1997}, which are a technical detail we will not discuss here):
\begin{align}
	\pi(ad[b]) &= \pi(a)\pi(d[b]) := a\cdot \Delta_D(b)\label{Eq_Hom_left}\\
		\pi(d[b]a) &= \pi(d[b])\pi(a) := \Delta_D(b)\cdot a \label{Eq_Hom_right}
\end{align}
$a,b\in A$. Indeed, it is quick to check that $\Delta_D: A\rightarrow \Omega_D^1 A$ satisfies the Leibniz rule, as required for a derivation:
\begin{align}
	\Delta_D[ab]&=[D,\pi_L(ab)]\nonumber\\
	&=[D,\pi_L(a)\pi_L(b)]\nonumber\\
	&=\pi_L(a)[D,\pi_L(b)]+[D,\pi_L(a)]\pi_L(b)\nonumber\\
	&= a\cdot \Delta_D[b] + \Delta_D [a]\cdot b \label{Eq_Leib_Dirac},
\end{align}
$a,b\in A$.

The challenge in the nonassociative setting is that even if one could sensibly create an analog of the universal calculus $\Omega_d^1 A$, two difficulties immediately arise: First,   in associative algebras commutators act as derivations. This ensures that $\Delta_D$ as defined in Eq.~\eqref{Eq._homomorphismD} acts as a derivation. Secondly, the standard construction of general 1-forms relies on the homomorphism property of associative representations outlined in Eqs.~\eqref{Eq_Hom_left} and \eqref{Eq_Hom_right}. Both of these properties fail in the case of nonassociative algebras: the inner derivations of  a nonassociative algebra $A$, denoted $Der(A)\in End(A)$, do not in general take the form of commutators, as shown in Eqs.~\eqref{Eq_Der_Jordan}, \eqref{Eq_Der_Lie},  \eqref{Eq_Der_Alterative}. Furthermore,  for nonassociative representations the homomorphism properties given in Eqs.~\eqref{Eq_Hom_left} and \eqref{Eq_Hom_right} generally fail. The challenge, which we overcome in this paper,  is to resurrect the spectral geometry construction by circumventing  each of these apparent obstructions.

\subsection{Derivation Bimodules}
\label{SSec_Derivation_Bimodules}

The challenges outlined in Sec.~\ref{SSec_Forms_As_Bimodules} motivate 
a novel definition for bimodules over nonassociative algebras. 
In particular,  the construction of a  bimodule with vector space  $M=A\otimes A$ over an algebra $A$, and with left and right products as given in Eqs.~\eqref{Eq_split_left_action} and \eqref{Eq_split_right_action} will generally necessitate dropping the requirement that $M$ should inherit the associative properties of $A$. Nevertheless $M$ should remain compatible with the inner derivations of $A$, and we are specifically interested in  cases in which there exists  non-trivial external derivations $\Delta:A\rightarrow M$.


\begin{definition}\label{def:bimodule2}
 Let $A$ be a (possibly nonassociative) algebra over a field $\mathbb{F}$. A \textbf{derivation bimodule} $M$ over $A$ is a square zero extension $B=A\oplus M$ in the sense of Def. \ref{def:sze},
equipped with a Lie algebra of derivations $Der(B)\subset End(B)$,  which satisfies the following compatibility conditions:
\begin{itemize}
	\item \textbf{Closure:} $Der(B)$ is closed under the Lie bracket and acts on $B$ via the Leibniz rule: $\delta(xy) = \delta(x)y+x\delta(y)$ for all $x,y\in B$, and $
	\delta\in Der(B)$. 	
	\item \textbf{Lifting Condition:} Every derivation $\delta\in Der(A)$ lifts to a derivation $\delta\in Der(B)$, such that  there exists an operator $\delta'\in End(M)$ satisfying
	\begin{align}
		\pi_L(\delta a) &= [\delta',\pi_L(a)]\label{lift1}\\
		\pi_R(\delta a) &= [\delta',\pi_R(a)]\label{lift2}
	\end{align}
	 for all $a\in A$, and where $\pi_L$ and $\pi_L$ denote the left and right actions of $A$ on $M$ respectively.
\end{itemize} 
\end{definition}

 To clarify  the above definition, it is helpful to examine derivation bimodules over associative algebras to investigate how they  compare with the standard notion of associative bimodules. To that end, consider a derivation bimodule $B = A\oplus M$, in which the  Lie algebra of Derivations  $Der(B)$ is defined to consist of  derivations of associative form as given in Eq.~\eqref{deralt2assoc}:
\begin{align}
\delta_c = L_c-R_c\label{derassoclie}
\end{align} 
 for all $c\in B$. For $\delta_c$ to act as a derivation means the Leibniz rule $\delta_c(ab) = \delta_c(a)b + a\delta_c(b)$ is satisfied for any $a,b,c\in B$. This   implies the following restrictions on the left and right $A$-action on $M$:
\begin{align}
\pi_L([a,b])-[\pi_L(a),\pi_L(b)]+[\pi_R(a),\pi_L(b)]&=0\label{leftassoc},\\
\pi_R([b,a])-[\pi_R(a),\pi_R(b)]+[\pi_R(a),\pi_L(b)]&=0,\label{rightassoc}\\
\pi_L(ba)-\pi_L(b)\pi_L(a)-\pi_R(ba)+\pi_R(a)\pi_R(b)-[\pi_R(a),\pi_L(b)]&=0,\label{weak3}
\end{align}
for $a,b\in A_F$. Taking combinations of the above identities, we can re-write the first two equations above in the following more convenient form:
\begin{align}
[\pi_R(a),\pi_L(b)]&=[\pi_L(a),\pi_R(b)],\label{weak1}\\		
\pi_L(ab)-\pi_L(a)\pi_L(b)&=\pi_R(ba)-\pi_R(a)\pi_R(b),\label{weak2}
\end{align}
for $a,b\in A$. The conditions given in Eqs.~\eqref{weak3}, \eqref{weak1}, and \eqref{weak2} are clearly compatible with, but weaker than the  usual defining conditions of an associative bimodule given in Eqs.~\eqref{Eq_Assoc_Module_1}, \eqref{Eq_Assoc_Module_2} and \eqref{Eq_Assoc_Module_3}. Nevertheless, they place stringent restrictions on the form that compatible representations can take. We provide two  examples exploring these identities in Appendix Section \ref{examplebimodules}.

\section{Split Alternative Bimodules And Gauge Invariance}
\label{Sec_Split_Alternative_Bimodules}

In this section, motivated by the explicit construction of associative universal 1-forms in Section~\ref{SSec_Forms_As_Bimodules},  we introduce split alternative bimodules over semi-simple, finite dimensional, alternative algebras.  We then explain why  if the spaces of forms in a discrete geometry, are assumed to be split alternative bimodules, then this gives rise to automorphism invariant Dirac operators and uncharged Higgs fields. This sets the context for  Section \ref{sec_example_geometry}, where we then show that it is indeed possible to construct models with Dirac operators that transform with respect to the symmetries of a discrete, alternative geometry, but this requires the introduction of a new class of derivation bimodule, which we name reconstituted alternative bimodules.

\subsection{Split Alternative Bimodules}
\label{SSec_Crossed_Derivations}

We would like to generalize the explicit construction of the space of universal 1-forms  $\Omega_d^1 A$ from the associative to the alternative setting. We are particularly interested in the case in which the algebra $A$ is finite dimensional and semi-simple, as such algebras coordinatize the internal spaces of gauge theories in the spectral geometric setting. To that end, consider the case of a unital, real, finite dimensional algebra, of the form $A=\bigoplus_i^n A_i$, where each factor $A_i$ is alternative. Such algebras correspond to discrete topologies, in which each factor $A_i$ encodes  a discrete `point'.  We construct an $A$-bimodule $M$ with vector space:
\begin{align}
	M = \bigoplus_{ij} A_i\otimes V^{ij}\otimes A_j, \label{Eq_split_mod}
\end{align}
where each $V^{ij}$ is a real vector space. Note that when $V^{ij}=\mathbb{R}$ for each $i,j$, we  have an isomorphism of real vector spaces $M \cong A \otimes A$, which corresponds to the standard space one begins with in the construction of universal 1-forms $\Omega_d^1 A$ in the associative setting as outlined in \ref{SSec_Forms_As_Bimodules}. We will call $M$ a split alternative bimodule over $A$ if it is equipped with following left and right products:
\begin{align}
	c_{(i)}\cdot (a\otimes v^{(jk)}\otimes b)&=\hat{\delta}^j_i(ca\otimes v^{(jk)}\otimes b),\label{Eq_left_alt}\\
	(a\otimes v^{(jk)}\otimes b)\cdot c_{(i)}&=\hat{\delta}^k_i(a\otimes v^{(jk)}\otimes bc),\label{Eq_right_alt}
\end{align}
where $c_{(i)} = (0,...c,...,0)\in A$, with a single entry in the $i^{th}$ column, and where $v^{(ij)}$ is an element of $V^{ij}$ such that $a\otimes v^{(jk)}\otimes b$ denotes an element of $A_j\otimes V^{jk}\otimes A_K$.   The symbol $\hat{\delta}^{ji}$ denotes the kronecker delta. We call such bimodules `split' alternative, because the left and right actions on $M$ are inherited seperately from the alternative product on the $A_i$, but are not jointly constrained. In particular, the  associator $[a,h,b] = (a\cdot h)\cdot b - a\cdot (h\cdot b)$ for $a,b\in A$, $h\in M$, fails the usual alternative identities $[a,h,b]=-[b,h,a]=-[a,b,h]=-[h,a,b]$.  One does, however, retain separate left and right alternative identities: $[a,b,h] = -[b,a,h]$ and  $[h,a,b] = -[h,b,a]$. 
One is therefore not able to relate associators of left acting elements to associators of right acting elements acting on $M$, and in this sense the bimodule is `split'. 

Split alternative bimodules satisfy the properties of derivation bimodules. In particular, each inner derivation $\delta\in Der(A)$ has a  lift to $M$ of the form:
\begin{align}
	\delta'(\omega_1\otimes v^{(ij)}\otimes \omega_2) =\delta(\omega_1)\otimes v^{(ij)}\otimes \omega_2 + \omega_1\otimes v^{(ij)}\otimes  \delta(\omega_2).\label{Eq_equivariant_der}
\end{align}
In fact, such lifts exist for split bimodules regardless of the associative properties of $A$, and it is indeed quick to check  that  they are compatible with the left and right actions of the algebra:
\begin{align}
	[\delta',\pi_L(a)](\omega_1\otimes\omega_2)&= \delta(a\omega_1)\otimes\omega_2 + a\omega_1\otimes \delta(\omega_2) - a\delta(\omega_1)\otimes\omega_2 - a\omega_1\otimes \delta(\omega_2)\nonumber\\
	&= \delta(a)\omega_1\otimes\omega_2= \pi_L(\delta(a))(\omega_1\otimes\omega_2)\\
	[\delta',\pi_R(a)](\omega_1\otimes\omega_2)&= \delta(\omega_1)\otimes\omega_2a + \omega_1\otimes \delta(\omega_2a) - \delta(\omega_1)\otimes\omega_2a - a\omega_1\otimes \delta(\omega_2)a\nonumber\\
	&= \omega_1\otimes\omega_2\delta(a)= \pi_R(\delta(a))(\omega_1\otimes\omega_2).
\end{align}
Furthermore,   split bimodules will always have `external' derivations $\Delta:A\rightarrow M$ as long as the algebra $A$ over which they are formed have a non-empty associative nucleus.  The associative nucleus of an algebra is given by  $Z_{ass}(A) = \{a\in A| [a,b,c]=[b,a,c]=[b,c,a] = 0, ~ \forall b,c\in A\}$. From any  element $a\in Z_{ass}(A)$, one can construct an inner derivation of associative form as given in Eq.~\eqref{deralt2assoc}, $\delta_a = L_a-R_a\in Der(A)$:
\begin{align}
	\delta_a(bc) &= a(bc) - (bc)a\nonumber\\
	&=(ab)c - b(ca)\nonumber\\
	&=[a,b]c + (ba)c - b(ca)\nonumber \\
	&= [a,b]c + b[a,c] = \delta_a(b)c + b\delta(c),
\end{align}
for all $b,c\in A$. When considering split bimodules, one can construct external   derivations of associative form $\Delta_a: A\rightarrow M $, from elements $a  \in Z_{ass}(B) \cap M$. In particular, consider the case in which one has the split bimodule $M=A\otimes A$ over a nonassociative algebra $A$ with unit $\mathbb{I}$. In this case one always has:
\begin{align}
	\mathbb{I}\otimes \mathbb{I} \in  Z_{ass}(B) \cap M,
\end{align}
which satisfies:
\begin{align}
	[a,b,\mathbb{I}\otimes \mathbb{I}] &= (ab)\mathbb{I}\otimes\mathbb{I} -   a(b\mathbb{I})\otimes\mathbb{I}=0\nonumber\\
	[a,\mathbb{I}\otimes \mathbb{I},b] &= a\otimes b -   a\otimes b=0\nonumber\\
	[\mathbb{I}\otimes \mathbb{I},a,b] &= \mathbb{I}\otimes (\mathbb{I}a)b -   \mathbb{I}\otimes\mathbb{I}(ab)=0
\end{align}
for all $a,b\in A$, and $\mathbb{I}\otimes \mathbb{I}\in M$. One therefore always has access to the external derivation:
\begin{align}
	\Delta_{\mathbb{I}\otimes \mathbb{I}}=L_{\mathbb{I}\otimes \mathbb{I}}-R_{\mathbb{I}\otimes \mathbb{I}}.\label{externalder}
\end{align}
 regardless of the associative properties of $A$. This is precisely the form of the 1st order derivation given in Eq.~\eqref{deri} when building the explicit construction of the associative universal calculus.

\subsection{Bimodule Homomorphisms beteween Split Alternative Bimodules}
\label{SSec_Split_Homomorphisms}

In the last section we isolated the external derivation $\Delta_{\mathbb{I}\otimes \mathbb{I}}$ given in Eq.~\eqref{externalder}, which will always exist for a split alternative bimodule of the form $M=A\otimes A$ defined over a unital alternative algebra $A$. In general, once an external derivation $\Delta:A\rightarrow M$ has been isolated for a bimodule $M$ defined over a (possibly nonassociative) algebra $A$,  one can then make use of what are known as bimodule homomorphisms to construct new derivations.   Given two bimodules $M$ and $N$ over an algebra $A$, a bimodule homomorphism between $M$ and $N$ is a linear map $\Phi:M\rightarrow N$, which commutes with the action of the algebra:
\begin{align}
	\pi^N_L(a)\phi(m) &= \phi(\pi_L^M(a)m),\label{Eq_module_mapL}\\
		\pi^N_R(a)\phi(m) &= \phi(\pi_R^M(a)m),\label{Eq_module_mapR}
\end{align}
for all $a\in A$, and $m\in M$. In the above expression $\pi_M:A\rightarrow End(M)$, and $\pi_N:A\rightarrow End(N)$ indicate the representations of $A$ on $M$ and $N$ respectively. In what follows, however, this notation will become cumbersome, and so we will instead rely on  context to indicate which bimodule the algebra elements are acting on. With this convention in place, Eqs. \eqref{Eq_module_mapL} and \eqref{Eq_module_mapR}  become:
\begin{align}
	a\cdot \phi(m) &= \phi(a\cdot m),\label{Eq_module_mapL2}\\
		\phi(m)\cdot a &= \phi(m\cdot a).\label{Eq_module_mapR2}
\end{align}
If $M$ is equipped with an external derivation $\Delta:A\rightarrow M$, and there exists a bimodule homomorphism $\Phi:M\rightarrow N$, then the composition $\Delta_\Phi =\Phi\circ \Delta$, will act as a derivation $\Delta_\Phi:A\rightarrow N$:
\begin{align}
	\Delta_\Phi[ab] &= 	\Phi(\Delta[ab])\nonumber\\
	& = \Phi(\Delta [a]\cdot b + a\cdot \Delta[b])\nonumber\\
	 & = \Phi(\Delta [a])\cdot b + a\cdot \Phi(\Delta[b]) = \Delta_\Phi[a]\cdot b+a\cdot \Delta_\Phi[b],
\end{align}
$a,b\in A$. A clear question, is to what extent bimodule homomorphisms can be used to extend the external derivation $\Delta_{\mathbb{I}\otimes \mathbb{I}}$ given in Eq.~\eqref{externalder} to   families of external derivations on more general split alternative bimodules.

 Given two split alternative bimodules 
of the form $M = \oplus_{ij} A_i\otimes V^{ij}\otimes A_j$ and $N = \oplus_{ij} A_i\otimes W^{ij}\otimes A_j$, over a unital, finite dimensional alternative algebra of the form $A =\oplus_i^n A_i$, a bimodule homomorphism $\Phi: M\rightarrow N$ will always take the form:
\begin{align}
	\Phi(a\otimes v^{(ij)}\otimes b) = a\otimes \Gamma w^{(ij)} \otimes b,\label{homotran}
\end{align} 
where $a\otimes v^{(ij)}\otimes b\in A_i\otimes V^{ij}\otimes A_j$, and 
 $\Gamma: V^{ij}\rightarrow W^{ij}$ is a linear map. To see this, notice that from the left one has:
\begin{align}
	a(b(\Phi(e_0\otimes v^{(ij)}\otimes c))) & = a(\Phi(b(e_0\otimes v^{(ij)}\otimes c)))\nonumber\\
	&=\sum_{ij}^n a(\Phi(b_{i}\otimes v^{(ij)}\otimes c))\nonumber\\
	&=\sum_{ij}^n \Phi(a_{i}b_{i}\otimes v^{(ij)}\otimes c)\label{LHScomp},
\end{align}
for $a= (a_1,...a_n),b=(b_1,...,b_n)\in A$, and where $e_0$ is the standard notation we will use for the identity element in any factor $A_i$ (as opposed to the identity element on $A=\oplus_i^n A_i$, which we denote $\mathbb{I}=\sum_i^n e_{(i)0}$). On the other hand one has:
\begin{align}
	(ab)(\Phi(e_0\otimes v^{(ij)}\otimes c)) & = \Phi((ab)(e_0\otimes v^{(ij)}\otimes c))\nonumber\\
	&= \Phi(a_ib_i\otimes v^{(ij)}\otimes c) = (ab)_{(i)}(\Phi(e_0\otimes v^{(ij)}\otimes c))\label{LHScomp2}.
\end{align}
 Comparing \eqref{LHScomp} and \eqref{LHScomp2}, then yields:
\begin{align}
	(ab)(\Phi(e_0\otimes v^{(ij)}\otimes c))=a(b(\Phi(e_0\otimes v^{(ij)}\otimes c))) = (ab)_{(i)}(\Phi(e_0\otimes v^{(ij)}\otimes c)), 
\end{align}
for all $a,b,c\in J_3(\mathbb{O})$, and $v\in V$. This implies that  $\Phi (e_0\otimes v^{(ij)}\otimes a)\propto (e_0\otimes \tilde{v}^{(ij)}\otimes \tilde{a})\in N$. A similar argument from the right also yields $\Phi (a\otimes v^{(ij)}\otimes e_0)\propto (\tilde{a}\otimes \tilde{v}^{(ij)}\otimes e_0)\in N$. We therefore have:
\begin{align}
	\Phi (e_0\otimes v^{(ij)}\otimes e_0) =  e_0 \otimes \Gamma w^{(ij)}\otimes e_0
\end{align}
where  $\Gamma$ is a linear map from $V^{ij}$ into  $W^{ij}$.  We then have:
\begin{align}
	\Phi (a\otimes v^{(ij)}\otimes b) &=a_{(i)}	\Phi (e_0\otimes v^{(ij)}\otimes e_0) b_{(j)}\nonumber\\
	&	=  a_{(i)}(e_0 \otimes \Gamma w^{(ij)}\otimes e_0)b_{(j)}=a \otimes \Gamma w^{(ij)}\otimes b\label{Eq_bimodhom}
\end{align}
This tells us that not only are bimodule homomorphisms between split alternative bimodules `trivial' in the sense that they only act on the vector spaces $V^{ij}$ and $W^{ij}$, but furthermore that they   respect the split bimodule decomposition. As a result the external derivations that one can form from the composition of the derivation given in Eq.~\eqref{externalder} with bimodule homomorphisms is extremely limited for split alternative bimodules, and this fact plays a key role when attempting to construct charged Higgs fields in physical models.

\subsection{Bimodule Homomorphisms And  Gauge Invariant 1st Order Derivations}
\label{SSec_Bimodule_Homomorphisms}

In section \ref{SSec_Forms_As_Bimodules} we introduced, an explicit construction of the universal 1-forms $\Omega_d^1 A$ over a unital, associative algebra with first order derivation map $\Delta_{\mathbb{I}\otimes\mathbb{I}}: A\rightarrow \Omega_d^1 A$ given as in Eq.~\eqref{leibder}. A key question is how this derivation map interacts with the symmetries of $A$ lifted to $\Omega_d^1 A$. The inner derivations of an associative algebra $\delta_a\in Der(A)$ can be lifted to a bimodule $M=A\otimes A$ in at least two distinct ways. The first derives from the associative form given in Eq.~\eqref{deralt2assoc}, and results from the associative product on  $B=A\otimes M$:
\begin{align}
	\delta'_a(\sum m\otimes n) = \sum am\otimes n - m\otimes na, 
\end{align}
$\sum m\otimes n\in \Omega_d^1 A$. The second form of lift is analogous to that  given in Eq.~\eqref{Eq_equivariant_der} for alternative algebras:
\begin{align}
	\delta'_a(\sum m\otimes n) = \sum \delta_a(m)\otimes n + m\otimes \delta_a(n).\label{Eq-liftnonas}
\end{align}
 It is  this second form of lift  that  we now analyse with respect to the external derivations $\Delta:A\rightarrow M$, as it is the form that generalizes immediately to split nonassociative bimodules. The  automorphisms $\alpha\in Aut(A)$  generated by $Der(A)$ are then lifted to $M$ as  $\alpha' = exp(\delta')$, with $\delta'$ given as in Eq.~\eqref{Eq-liftnonas}:
 \begin{align}
 	\alpha'(\omega_1\otimes\omega_2) = \alpha(\omega_1)\otimes\alpha(\omega_2). \label{Eq_lifted_aut}
 \end{align}
   The first order derivation map given in Eq.~\eqref{deri} is blind to the derivations lifted as in \eqref{Eq-liftnonas} to $\Omega_d^1A$:
\begin{align}
	\delta'(\Delta_{\mathbb{I}\otimes \mathbb{I}}[a])&= \delta'(a\otimes \mathbb{I} - \mathbb{I}\otimes a)\nonumber\\
	&=\delta(a)\otimes \mathbb{I} - \mathbb{I}\otimes \delta(a) = \Delta_{\mathbb{I}\otimes \mathbb{I}}[\delta(a)],
\end{align} 
$a\in A$, where the second equality  holds as  the identity element  $\mathbb{I}\in A$ is annihilated by derivations.  As such the operator $\Delta:A\rightarrow M$ commutes with the lifted automorphisms  generated by the Lie algebra $Der(A)$:
\begin{align}
	\alpha'\Delta_{\mathbb{I}\otimes \mathbb{I}} = \Delta_{\mathbb{I}\otimes \mathbb{I}}\alpha, 
\end{align} 

When constructing physical models, the 1st order derivations $\Delta:A\rightarrow M$ we are interested in are related to Dirac operators  as shown in Eq.~\eqref{Eq._homomorphismD}. For this reason, what we would  ultimately like are not first order derivations which are invariant with respect to the symmetries of the representation, but rather which transform with respect to the lifted automorphisms. This is because  Dirac operators in spectral geometry encode gauge and Higgs fields in physical models, which transform with respect to the gauge symmetries. To achieve covariance we can take advantage of bimodule homomorphisms to construct new derivation maps $\Delta_\Phi = \Phi\circ \Delta$.  Derivations $\Delta_\Phi$ constructed in this way will not in general remain  invariant with respect to the symmetries lifted to $M$ and $N$. In particular, starting with a gauge invariant external derivation $\Delta_{\mathbb{I}\otimes \mathbb{I}}:A\rightarrow M$ from Eq.~\eqref{deri}, and composing with a bimodule Homomorphism $\Phi:M\rightarrow N$, one has: 
\begin{align}
	\Delta_\Phi[\alpha(a)]&=	\Phi(\Delta_{\mathbb{I}\otimes \mathbb{I}}[\alpha(a)])\nonumber\\
	 &= \Phi(\alpha_M'(\Delta_{\mathbb{I}\otimes \mathbb{I}}[a])) \nonumber\\
	 &= \alpha_N'\Phi'(\Delta_{\mathbb{I}\otimes \mathbb{I}}[a]) = \alpha_N'\Delta_{\Phi'}[a]
\end{align}
where $\Phi'= (\alpha_N')^{-1}\circ\Phi\circ \alpha_M'$, and the subscripts indicate which module the lifted automorphism $\alpha'$ acts on.  As such, the derivative operator $\Delta_\Phi$ will in general transform:
\begin{align}
	\Delta_{\Phi}\rightarrow  \Delta_{\Phi'}
\end{align}
with respect to the automorphisms of the algebra, unless the homomorphism map $\Phi:M\rightarrow N$ itself commutes with the symmetries of the representation. Consider for example bimodule maps $\Phi:\Omega_d^1 A\rightarrow A\otimes A$ of the form:
\begin{align}
	\Phi(\omega_1\otimes \omega_2)\rightarrow \sum \omega_1b\otimes c\omega_2 \label{Eq_homo_assoc}
\end{align}
with the sum taken over the elements $b,c\in A$. One always has access access to homomorphisms of this form in the associative setting. Such maps produce new derivation operators $\Delta_\Phi:A\rightarrow A\otimes A$ of the form:
\begin{align}
	\Delta_\Phi[a] = \sum ab\otimes c - b\otimes ca,\label{Eq_general_order_1}
\end{align}
$a,b,c\in A$. Note that it should be no surprise that it is possible to construct whole families of derivations in this way in the associative setting.  In the associative setting, when considering an associative bimodule $M=A\otimes A$, any element $b\otimes c\in M$ sits in $Z_{ass}(B)$, and so provides a first order inner derivation. As such, $\Delta_\Phi$ in Eq.~\eqref{Eq_general_order_1} is nothing other than a derivation of associative form $\Delta_\Phi=\sum R_{b\otimes c}-L_{b\otimes c}$. 
Importantly, the new derivation that we have generated $\Delta_\Phi$,  transforms under the lifted automorphisms as follows:
\begin{align}
	\Phi(\Delta [\alpha(a)]) &= \sum (\alpha(a)b\otimes c - b\otimes c\alpha(a))\nonumber\\
	&= \sum (\alpha(a\alpha^{-1}(b)\otimes \alpha(\alpha^{-1}(c)) - \alpha(\alpha^{-1}(b))\otimes \alpha(\alpha^{-1}(c)a))\nonumber\\
	&=	\alpha'  \sum (a\alpha^{-1}(b)\otimes \alpha^{-1}(c) - \alpha^{-1}(b)\otimes \alpha^{-1}(c)a) = \alpha' \Delta_{\Phi'}[a]
\end{align}
where the transformed homomorphism $\Phi'$ is given by:
\begin{align}
	\Phi'(\omega_1\otimes \omega_2)\rightarrow \sum \omega_1\alpha^{-1}(b)\otimes \alpha^{-1}(c)\omega_2. 
\end{align}
We see that in the associative setting, composing an automorphism invariant external derivation $\Delta_{\mathbb{I}\otimes \mathbb{I}}:A\rightarrow M$, with bimodule homomorphisms $\Phi:M\rightarrow N$ will in general result in derivation maps   $\Delta_\Phi:A \rightarrow N$, which transform with respect to the lifted automorphisms of the algebra $A$.  As we will now see, the situation differs considerably in the alternative setting, where bimodule homomorphisms take the restricted form given in Eq.~\eqref{homotran}.

Consider a split alternative bimodule $M$ of the form given in Eq.~\eqref{Eq_split_mod}, defined over a finite dimensional,  unital, alternative  algebra $A=\bigoplus^n_i A_i$ with left and right actions given as in Eqs.~\eqref{Eq_left_alt} and \eqref{Eq_right_alt}. Furthermore, consider the case in which the bimodule $B=A\oplus M$ is equipped with an external automorphism invariant derivation  $\Delta:A\rightarrow M$.
Our goal is to explore the interaction between the continuous lifted  automorphisms of $A$ and 1st order derivations $\Phi\circ\Delta:A\rightarrow M$, where $\Phi:M\rightarrow M$ is a bimodule homomorphism, 
 to see whether we can construct  external derivations that transforms under the symmetries of the bimodule. 
  Unfortunately all  homomorphisms between split alternative bimodules take the form given in Eq.~\eqref{homotran}, which commute with lifted automorphisms:
\begin{align}
	\Phi(\alpha_M'(a\otimes v^{(ij)}\otimes b)) &=	\Phi(\alpha(a)\otimes v^{(ij)}\otimes \alpha(b)) \nonumber\\
	&= \alpha(a)\otimes \Gamma v^{(ij)}\otimes \alpha(b)\nonumber\\
	&= \alpha_N'(a\otimes \Gamma v^{(ij)}\otimes b) = \alpha_N'(\Phi(a\otimes  v^{(ij)}\otimes b)).
\end{align} 
 This means that all derivations constructed through composition with bimodule homomorphisms $\Delta_\Phi = \Phi\circ \Delta_{\mathbb{I}\otimes \mathbb{I}}$, will remain automorphism invariant in the split alternative setting.
 
\subsection{Split Alternative Bimodules over Discrete Octonion Algebras And Automorphism  Invariant 1-st Order Derivations}
\label{Sec_oct_invariance}

In Section \ref{SSec_Bimodule_Homomorphisms} we showed that unlike in the associative  setting, starting with an automorphism invariant external  derivation, one is not able to construct  covariant derivations through composition with  bimodule homomorphisms between split alternative bimodules. This, however, does not preclude the existance of gauge covariant derivations that are not related by bimodule homomorphisms. For instance, one might have access to derivation compatible maps $\Phi:M\rightarrow N$:
\begin{align}
	\Phi\circ \Delta(ab)= 	\Phi\circ \Delta(b)\cdot b+	a\cdot \Phi\circ \Delta(b),\label{dericompat}
\end{align}
$a,b\in A$, which nevertheless fail to satisfy the properties of bimodule homomorphisms. In this section we restrict attention to discrete alternative algebras of the form $A = \bigoplus_i^n A^i$, where each factor $A_i$ is a copy of the octonions $\mathbb{O}$. In this case we know the structure constants of the algebra, and so by direct computation   we can show  that   automorphism covariant external derivations don't exist.

Consider the algebra  $A = \bigoplus_i^n A^i$, where each $A_i$ is a copy of the octonions $\mathbb{O}$. In this case the vector space of any split alternative bimodule $M$ over $A$ can be expressed as:
\begin{align}
	M  \cong \mathbb{O}\otimes V\otimes \mathbb{O}, 
\end{align}
where $V = \bigoplus_{ij} V^{ij}$, and where the split alternative product given in Eqs.~\eqref{Eq_left_alt} and \eqref{Eq_right_alt} are re-expressed as:
\begin{align}
	a\cdot(h\otimes v \otimes k) &= \sum_{ij}^n a_i h\otimes P^{ij}(v)\otimes h,\\
		(h\otimes v \otimes k)\cdot a &= \sum_{ij}^n a_i h\otimes P^{ij}(v)\otimes ha_j,
\end{align}
where $a = (a_1,..., a_n)\in A$, and $P^{ij}$ projects onto the subspace $P^{ij}(V) = V^{ij}$. We now restrict attention  to the case in which each $V^{ij}=\mathbb{R}$ such that $M\cong A\otimes A$. We define the convenient basis notation
 $e^{(a)I}$ corresponding to the $e^I$ basis element of the $a^{th}$ factor of the coordinate algebra:
\begin{align}
	e^{(a)I} := (0, \dots, e^I, \dots, 0) \in A,
\end{align}
and similary the following basis elements for $M$
\begin{align}
	e^{I}\otimes^{(ab)} e^I := (0,...,e^{i}\otimes e^j,...,0)\in  M\label{projectedvect}
\end{align}
where the $e^{I}\otimes e^J\in \mathbb{O}\otimes \mathbb{R}\otimes \mathbb{O}$ are the basis elements of $\mathbb{O}\otimes V^{ab}\otimes \mathbb{O}$. Note that we use capital indices when referring to the full basis of the octonions including the identity element $I,J = 0,...,7$. When referring only to the imaginary basis elements we will index with lower case letters $i,j=1,...,7$. 
 With this basis established the most general linear operator $\Delta:A\rightarrow A\otimes A$ is of the form:
\begin{align}
	\Delta[e^{(a)i}] = \Delta^{(a)i}_{(bc)jk} e^j\otimes^{bc}e^k,
\end{align}
where the $\Delta^{(a)i}_{(bc)jk}$ are real coefficients. For $\Delta$ to act as a derivation it must further satisfy the Leibniz condition given in  Eq.\eqref{leibder}:  $\Delta[e^{(a)i}e^{(b)j}] = \Delta[e^{(a)i}]\cdot e^{(b)j}+e^{(a)i}\cdot \Delta[e^{(b)j}]$. This implies the following relations, which must be satisfied by the $\Delta^{(a)i}_{(bc)jk}$ coefficients:
\begin{align}
\hat{\delta}^{(ab)}f^{ij}_k	
\Delta^{(b)k}_{(cd)st} 
 =\hat{\delta}^{(bd)}f^{nj}_t
\Delta^{(a)i}_{(cd)sn} +\hat{\delta}^{(ac)}f^{im}_s
\Delta^{(b)j}_{(cd)mt} \label{Eq_leib_to_sat}
\end{align}
where the $f^{ij}_k$ are the structure constants that define the product on each copy of $\mathbb{O}$, and where Einstein's summation convention is not being used with  bracketed indices.  While this expression looks cumbersome, it  decomposes neatly into a number of manageable  cases. In particular, if we consider the case in which $c\neq a=b\neq d$, then Eq.~\eqref{Eq_leib_to_sat}  becomes:
\begin{align}
	f_k^{ij}\Delta^{(b)k}_{(cd)mn} =0,& &\text{for }b\neq c\text{ and }b\neq d,
\end{align}
which tells us immediately that:
\begin{align}
	\Delta^{(b)}_{(cd)} &=0, & \text{for }&b\neq c\text{ and }b\neq d,
\end{align}
where $\Delta^{(b)}_{(cd)}$ correspond to the components of $\Delta$ that map from basis elements $e^{(b)i}$ into the subspace of $M=\mathbb{O}\otimes V\otimes \mathbb{O}$ isolated by the  $P^{bc}$ projector  on $V$. We therefore only need to worry about 
the components of $\Delta$ that map from basis elements $e^{(a)i}$ into the subspace of $M$ isolated by projectors of the form $P^{ab}$ and $P^{ba}$  on $V$.  Starting with the  `diagonal' components of $\Delta$ that map basis elements $e^{(a)i}$ into the subspace of $M$ isolated by $P^{aa}$, we set $a=b=c=d$ in Eq.~\eqref{Eq_leib_to_sat}, which  yields:
\begin{align}
	f_k^{ij}\Delta^{(a)k}_{(aa)st}  =  f^{nj}_t\Delta^{(a)i}_{(aa)sn}+
	f^{im}_s	  \Delta^{(a)j}_{(aa)mt} 
\end{align}
This equation can be solved in Mathematica, and yields:
\begin{align}
	\Delta^{(a)}_{(aa)}[e^{(a)i}] &= \kappa_{(aa)}(e^i\otimes^{(aa)} e^0 - e^0\otimes^{(aa)} e^i), & \kappa_{(aa)}\in \mathbb{R}.\label{aadisc}
\end{align}
We can similarly isolate the components $\Delta^{(a)}_{(ab)}$ and $\Delta^{(a)}_{(ba)}$ of $\Delta$ by considering the cases $c=a\neq b=d$,  $a=b=c\neq d$, and $a=b=d\neq c$ in Eq.~\eqref{Eq_leib_to_sat}. This yields the following three  coupled equations:
\begin{align}
	f^{kj}_t\Delta^{(e)i}_{(ef)sk}+
	f^{ik}_s	  \Delta^{(f)j}_{(ef)kt} &=0\\
	f_k^{ij}\Delta^{(e)k}_{(ef)st} - 
	f^{ik}_s	  \Delta^{(e)j}_{(ef)kt}&=0 \\
	f_k^{ij}\Delta^{(f)k}_{(ef)st}  -  f^{kj}_t\Delta^{(f)i}_{(ef)sk}&=0
\end{align}
Solving these equations through an explicit  computation performed using Mathematica, we find that derivations are restricted to be of the form:
\begin{align}
	\Delta^{(c)}_{(ab)}[e^{(c)i}] &= \kappa_{(ab)}(\delta^{(ca)}e^i\otimes^{(ab)} e^0-\delta^{(bc)}e^0\otimes^{(ab)} e^i), & \kappa_{ab}&\in\mathbb{R}.\label{abdisc}
\end{align}
Combining Eq.~\eqref{aadisc} and \eqref{abdisc}, we therefore have in general
\begin{align}
	\Delta_\kappa[e^{(a)i}] = \sum_c  (\kappa_{(ac)} e^i\otimes^{(ac)}  e^0 - \kappa_{(ca)}e^0\otimes^{(ca)} e^i ).\label{Eq_split_alt_der}
\end{align}
This is the same result as that obtained for split Jordan bimodules over the exceptional Jordan algebra in 
\cite{farnsworth2025npointexceptionaluniverse}. We see that up to the coefficients $\kappa_{(ab)}$ there is a single unique form that derivations can take from $A$ into $A\otimes A$. Furthermore, from Eq.~\eqref{homotran} we see that the freedom that does exist stems from  bimodule morphisms $\Phi_\kappa:M\rightarrow M$ of the form $	\Phi_\kappa(a\otimes^{(ab)}\otimes b) = 	\kappa_{ab}(a\otimes^{(ab)} b)$. The standard derivation $\Delta_{\mathbb{I}\otimes \mathbb{I}}$ corresponds to the case in which all coefficients $\kappa_{ab}$ is set to 1. 

By analogy with the  associative construction of the space of universal 1-forms, we denote   the submodule of $M=A\otimes A$ generated as a bimodule by elements of the form $\Delta_{\mathbb{I}\otimes \mathbb{I}}[a]$ by $\Omega_\Delta^1 A$.  
It turns out, however, that  unlike in the associative case, such elements generate all of  $M= A\otimes A$ as a bimodule. We provide a proof of this fact in Appendix Section~\ref{sec_identify}. Furthermore, it turns out that the space of 1-forms $\Omega_\Delta^1 A$ is universal in the following sense: Given any split alternative bimodule of the form $M =\mathbb{O}\otimes V\otimes \mathbb{O}$, defined over the octonion algebra $A=\oplus_1^n \mathbb{O}$,
with left and right actions defined as in Eqs.~\eqref{Eq_left_alt} and \eqref{Eq_right_alt}, and  with derivation $\Delta:A\rightarrow M$, then there exists a unique bimodule homomorphism $\Phi:\Omega_\Delta^1 A \rightarrow M$ such that  $\Delta = \Phi\circ \Delta_{\mathbb{I}\otimes \mathbb{I}}$. In particular, following section \ref{SSec_Split_Homomorphisms}, if $\Phi$ is a module homomorphism then it will be of the form:
\begin{align}
	\Phi(\Delta[e^{(a)i}])  = \sum_c  ( e^i\otimes v^{(ac)}\otimes  e^0 - e^0\otimes v^{(ca)}\otimes e^i )
\end{align}
$a\in A$, and where the $v^{ab}\in V^{ab}$ are real vectors associated to the module map $\Phi$. It is clear that the $v^{(ac)}$ define $\Phi$ uniquely. Furthermore, it is quick to show that $\Phi(\Delta_{\mathbb{I}\otimes \mathbb{I}}[ab])=\Delta[ab]=a\Delta[b]+\Delta[a]b=\phi(a\Delta_{\mathbb{I}\otimes \mathbb{I}}[b]+\Delta_{\mathbb{I}\otimes \mathbb{I}}[a]b)$, and therefore that $\Phi$ respects the Lebniz relation of $\Delta_{\mathbb{I}\otimes \mathbb{I}}$ on $\Omega_\Delta^1 A$. For this reason we  identify $\Omega_\Delta^1 A$ as the split alternative universal 1-forms $\Omega_d^1 A$ over the octonion algebra $A=\oplus^n_1 \mathbb{O}$, with the differential $d:A\rightarrow \Omega_d^1 A$  given by:
\begin{align}
	d[a]&= \Delta_{\mathbb{I}\otimes \mathbb{I}}[a], & a&\in A.
\end{align}

\section{Spectral Geometry And The Construction Of Charged Dirac Operators}
\label{sec_example_geometry}

In Section \ref{Sec_Split_Alternative_Bimodules} we showed how the properties of split alternative bimodules appear to rule out the possiblity of constructing discrete spectral geometries with Dirac operators that transform with respect to the symmetries of the coordinate algebra, and  therefore rule out interesting physical models with $G_2$ symmetry and charged Higgs fields. As we will show by explicit example in this section, however, it \textit{is} in fact possible to build
models with covariant 1st order derivations by sculpting the properties of the bimodule of 1-forms to accommodate the automorphism covariant derivations of interest. This is the physical motivation for the introduction of derivation bimodules.
We first start in Section \ref{Sec_oct_model} by  introducing an example geometry corresponding to the internal space of a $G_2\times G_2$ gauge theory. We outline the coordinate algebra, Hilbert space representation, and general form for the Dirac operator of an alternative spectral triple. In section \ref{sec_uncharged} we then focus on the case in which the space of 1-forms $\Omega_D^1 A$ is described as a split alternative bimodule over the coordinate algebra, and show that indeed we end up with a Dirac operator that commutes with the dervations of the continuous symmetries of the model. In Section \ref{sec_charged} we then make use of the specialization rule given in Eq.~\eqref{specialization}   to derive a new form of `reconstituted' alternative bimodule. We then show  that it is possible to build charged Dirac operators by constructing the space of 1-forms as a `reconstituted' alternative bimodule, which  accommodates automorphism covariant external derivations.

\subsection{The Internal Spetral Data $T_F = (A_F,D_F,H_F)$ of a $G_2\times G_2$ Gauge Theory.}
\label{Sec_oct_model}

 The spectral geometries relevant to gauge theories typically take the form of product spaces:
\begin{align}
	T = T_c\times T_F,
\end{align}
where  $T_c = (C^\infty(M), L^2(M,S),\nabla^S)$ is a continuous, commutative spectral triple that  encodes information about the external space including  the 4D manifold and metric data, and $T_F =(A_F,H_F,D_F)$ is a discrete, finite dimensional spectral triple that encodes the `internal' degrees of freedom that describe particle species and representations.
Our focus  will be on constructing the  internal  data $T_F$ of a $G_2\times G_2$ gauge theory,  relegating the construction of the total space $T=T_c\times T_F$ corresponding to a complete gauge theory to Appendix Secion~\ref{Sec_quat_model}. We take this approach as the internal space $T_F$ contains all of the  novel elements of the current paper, while the general framework for constructing gauge theories, given a set of finite data $T_F$, is discussed exhaustively elsewhere~\cite{Chamseddine:2007oz,Dungen}. 
Our task, then,  is to construct the spectral data $T_F = (A_F,D_F,H_F)$ corresponding to the internal space of a nonassociative gauge theory. We select the internal  coordinate algebra $A_F = \mathbb{O}\oplus \mathbb{O}$, represented on a copy of its own vector space $H_F = A_F$ with the following left and right actions:
\begin{align}
	\pi_L(a) &= \begin{pmatrix}
		L_{a_1} & 0\\
		0&L_{a_2}\\
	\end{pmatrix}, & 	\pi_R(a) &= \begin{pmatrix}
		R_{a_1} & 0\\
		0&R_{a_2}\\
	\end{pmatrix},\label{leftrightoctaction}
\end{align}
for $a=(a_1,a_2)\in A_F$, and where $L_ab = ab$ and $R_ab = ba$ for $a,b\in \mathbb{O}$. This representation satisfies all of the properties of a nonassociative bimodule in the traditional sense outlined in Def.~\ref{def:bimodule}, such that $B_F = A_F\oplus H_F$ is itself an alternative algebra, and such that the continuous symmetries of $A_F$ lift readily to $H_F$. The following identities can be derived from the alternative properties of the representation:
\begin{align}
	\pi_L(ab)&=\pi_L(a)\pi_L(b) +[\pi_L(a),\pi_R(b)],\label{eq_alt_1}\\
	\pi_R(ab)&=\pi_R(b)\pi_R(a)+[\pi_R(b),\pi_L(a)],\label{eq_alt_x}\\
	[\pi_L(a),\pi_R(b)]&=[\pi_R(a),\pi_L(b)],\label{eq_alt_2}
\end{align}
$a,b\in A$. Equations \eqref{eq_alt_1} and \eqref{eq_alt_x} are specialization rules  as in Eq.~\eqref{specialization}. 
Furthermore, the representation space  $H_F = \mathbb{R}^{16}$ is a real Hilbert space, equipped with a  natural inner product that derives from the octonionic norm, and which we discuss in Appendix~\eqref{Sec_hilmb}.

Notice that although we have introduced both left and right actions of $A_F$ on $H_F$ in Eq.~\eqref{leftrightoctaction}, these are in fact not independent from one another. To see why, notice that unlike with an associative algera, for a nonassociative algebra one has in general $L_aL_b\neq L_{ab}$. For the octonions, it turns out that any linear operator on $\mathbb{O}$ can be created from sums of chains of left acting algebra elements, and this includes  $R_a$ for any $a\in\mathbb{O}$~\cite{Furey_2018}. For this reason the  right action can be thought of as  useful and compact notation, which will simplify the discussion that follows. Furthermore, the octonions are an involutive algebra, and the involution $\ast$ can be used to express the right action  of the octonions on themselves in terms of the left action:
\begin{align}
	R_ab = (a^\ast b^\ast)^\ast = (L_{a^*} b^\ast)^\ast,
\end{align}
$a\in \mathbb{O}$. If we introduce the notation $j(a) :=a^*$, then the right action $\pi_R:A_F\rightarrow End(H_F)$ can be expressed as:
\begin{align}
	\pi_R(a) = J_F\pi_L(a)^\ast J_F,\label{eq_right_action}
\end{align}
where the involution on the RHS is given by the transpose, and 
\begin{align}
	J_F := \begin{pmatrix}
		j&0\\
		0& j
	\end{pmatrix}.\label{InvolJ}
\end{align}
Because our algebra $A_F=\mathbb{O}\oplus \mathbb{O}$ is not simple, the 
internal octonionic geometry $T_F$ corresponds to a discrete geometry. The two octonion factors describe  discrete points, which are initially decoupled from one another with respect to the algebra product and the representation. In particular, following Eq.~\eqref{Eq_Der_Alterative}, the continuous symmetries of the theory are generated by inner derivation elements of the form~\cite{Schafer}:
\begin{align}
	\delta'_{ab} = [\pi_L(a),\pi_L(b)]+[\pi_L(a),\pi_R(b)]+[\pi_R(a),\pi_R(b)]\label{inner_der_alt}
\end{align}
for $a,b\in A_F$. The operators $\delta_{ab}'\in End(H_F)$ lift the generators $\delta_{ab}\in Der(A_F)$  of the same alternative form to $H_F$. These generators are block diagonal, and therefore act `internal' to each point of the geometry, and do not cause any mixing between them.

The discrete points of the geometry can be connected to each-other by introducing an appropriate space of 1-forms, which are also represented on $H_F$. This is done by introducing a block off-diagonal, Hermitian  Dirac operator $D_F:H_F\rightarrow H_F$, of the following general form:
\begin{align}
	D_F &= M_{IJ}\begin{pmatrix}
		0 & e_I\otimes e_J^\ast\\
		e_J\otimes e_I^\ast&0\\
	\end{pmatrix},\label{altDf}
\end{align}
where the `$\ast$' indicates the dual with respect to the Hilbert space $H_F$. Following Eq.~\eqref{Eq._homomorphismD} The Dirac operator is required to act as a  derivation with respect to the representation of $A_F$ on $H_F$:
\begin{align}
	D_F\pi_L(a)h &= \Delta_D(a)h + \pi_L(a)D_Fh\nonumber\\
	\rightarrow \Delta_D(a) &= [D_F,\pi_L(a)], \label{Exterior_D}
\end{align} 
$a\in A_F$, $h\in H_F$. Just as $\pi_L(a)$ provides a representation of $a\in A$ in $End(H)$, the idea is that the  operator $\Delta_D: A_F\rightarrow  \Omega_D^1A\subset End(H)$ provides a representation of the universal  `exact' form $d[a]\in \Omega_d^1 A$ in $End(H_F)$. For now we are agnostic about the properties of $\Omega_d^1 A$, and do not enforce that it must take the form of a   split alternative bimodule  over $A_F$ (see Sec.~\ref{Sec_oct_invariance}). 
Instead, we simply ask that $\Delta_D$ maps from $A_F$ to the space of 1-forms $\Omega_D^1 A$, which is generated by the elements $\Delta_D[a]$ as a bi-module over $A_F$, and which simultaneously provides a representation of $\Omega_d^1A_F$  on $H_F$, whatever form it may take. 
As a Derivation, $\Delta_D$ must satisfy a Leibniz rule of the following form when acting on elements of $A_F$:
\begin{align}
	\Delta_D(ab) = \Delta_D(a)\cdot b + a\cdot\Delta_D(b)\label{Deralt}
\end{align}
for some appropriate left and right actions $\cdot:A_F\times  \Omega_D^1A_F\rightarrow \Omega_D^1A_F$, $\cdot: \Omega_D^1A_F\times A_F\rightarrow \Omega_D^1A_F$. Ensuring that the above Leibniz rule holds   will place severe restriction on the form of the Dirac operator $D$, and on the left and right bimodule actions. To understand what restrictions the Leibniz rule might impose, we can  make use of the alternative identity given in Eqs.~\eqref{eq_alt_1}, together with the form of $\Delta_D$ given in Eq.~\eqref{Exterior_D} to write
\begin{align}
	\Delta_D(ab)&=	[D_F,\pi_L(a)\pi_L(b)+[\pi_L(a),\pi_R(b)]] \nonumber\\
	&= \Delta_D(a)\pi_L(b)+\pi_L(a)\Delta_D(b) +[D_F,[\pi_L(a),\pi_R(b)]],\label{leibrulealt}
\end{align}
$a,b\in A_F$. Comparing  Eqs.~\eqref{Deralt} and \eqref{leibrulealt}, we see that if $D_F$ is to act as a derivation, then the left and right actions   that define the module $\Omega_D^1 A_F\in End(H_F)$, will likely be given by something other than the standard operator product. We will explore the restrictions on $\Omega_D^1 A_F$ that arise from the external derivation $\Delta_D$ in this way  in sections \ref{sec_uncharged} and \ref{sec_charged}. For now we point out that  if  $\Omega_D^1 A$ is to act as a derivation bimodule over the $A_F$, then it must also be compatible with the inner derivations of $A_F$ lifted to  $\Omega_D^1 A$:
\begin{align}
	\delta_{ab}(c\cdot \omega) = \delta_{ab}(c)\cdot \omega + c\cdot 	\delta_{ab}(\omega)\label{innerleftL}\\
	\delta_{ab}( \omega\cdot c) =  	\delta_{ab}(\omega)\cdot c +  \omega\cdot \delta_{ab}(c)\label{innerleftR}
\end{align}
$c\in A_F$, $\omega\in \Omega_D^1 A_F$, and for inner derivations $\delta_{ab}\in Der(A_F)$. We can determine the form that  derivations take when lifted to $\Omega_D^1 A_F$ by noticing that because  $A\oplus \Omega_d^1 A$, is represented on $H$, the derivations lifted to $\Omega_D^1 A$  can be expressed in terms of the lift to $H_F$. Specifically, given a 1-form element $\omega\in\Omega_D^1 A_F$,  inner derivations of the form given in Eq.~\eqref{inner_der_alt} act as:
\begin{align}
	\delta_{ab}'\omega h = \delta_{ab}(\omega)h + \omega \delta_{ab}'h, 
\end{align}
$\omega\in\Omega_D^1 A_F$, $h\in H_F$. This tells us the action of elements $\delta_{ab}\in Der(A_F)$ are `lifted' to $\Omega_D^1 A_F$ with the following action:
\begin{align}
	\delta_{ab}(\omega) = [\delta_{ab}',\omega],\label{eq_action_on_1forms}
\end{align}
$\omega\in \Omega_D^1 A_F$. Equations \eqref{innerleftL} and \eqref{innerleftR} therefore become
\begin{align}
	[\delta'_{ab},c\cdot \omega] &= \delta_{ab}(c)\cdot \omega + c\cdot 	[\delta_{ab}',\omega],\label{innerleftL2}\\
	[ \delta_{ab}', \omega\cdot c] &=  	[\delta_{ab}',\omega]\cdot c +  \omega\cdot \delta_{ab}(c).\label{innerleftR2}
\end{align}
What is clear, is that it is the lifted inner derivations $\delta_{ab}'$, and the external derivation $\Delta_D$ constructed in terms of $D_F$, which constrain the form of the bimodule $\Omega_D^1 A_F$ over $A_F$. To explore these constraints  we now consider two different kinds of restriction on the finite Dirac operator. Different choices of $D_F$ will establish different sets of Derivations $Der(B)$, and therefore establish different properties for  $\Omega_D^1 A_F$ as a derivation bimodule over $A_F$.

\subsection{The Construction of Uncharged Higgs fields.}
\label{sec_uncharged}

A special case for the finite Dirac operator occurs when we set  $M_{ij}  = M_{0j} = M_{i0}= 0$ for all $i,j = 1,...7$ in Eq.~\eqref{altDf}:
\begin{align}
	D_F = M_{00}\begin{pmatrix}
		0 & e_0\otimes e_0^\ast\\
		e_0\otimes e_0^\ast&0\\
	\end{pmatrix}.\label{unchargedD}
\end{align}
The identity element
$e_0\in\mathbb{O}$ is  invariant under the inner automorphisms of the octonions, and as such, for this restricted form of the Dirac operator
and for inner derivation elements of the form given in Eq.~\eqref{inner_der_alt},  
it is easy to show:
\begin{align}
	[D_F,\delta_{ab}'] = 0,
\end{align}
for all $a,b\in A_F$. This means that the Dirac operator is completely blind to the continuous symmetries of the geometry. If we were to construct a gauge theory as a product geometry $T=T_c\times T_F$, then this Dirac operator would give  rise to a single, real, uncharged Higgs field corresponding to the single free parameter in the internal space $M_{00}$. This field is similar to the uncharged  $\sigma$ field that arises in the noncommutative standard model~\cite{Chamseddine:2012fk}, and to the uncharged Higgs fields in our earlier $F_4^n$ models based on exceptional Jordan algebras~\cite{farnsworth2025npointexceptionaluniverse}.

If we assume the `uncharged' form for the Dirac operator given in Eq.~\eqref{unchargedD},  then the Leibniz rule given in Eq.~\eqref{leibrulealt} reduces to the following form: 
\begin{align}
	\Delta_D(ab)&=	[D_F,\pi_L(a)\pi_L(b)+[\pi_L(a),\pi_R(b)]] \nonumber\\
	&= \Delta_D(a)\pi_L(b)+\pi_L(a)\Delta_D(b), \label{leibuncharged}
\end{align}
$a,b\in A_F$. To understand why, notice that because the identity element $e_0$ is in the associative nucleus of the octonion algebra, this implies $[R_a,L_b]e_0 = [b,e_0,a]=0 $ for all $a,b\in \mathbb{O}$. We therefore have $[D_F,[\pi_L(a),\pi_R(b)]]=0$, whenever the Dirac operator is constructed from identity elements as in Eq.~\eqref{unchargedD}.  Comparing Eqs.~\eqref{leibuncharged} and Eq.~\eqref{Deralt}, we then see that the uncharged Dirac operator is compatible with the following left and right  bimodule actions:
\begin{align}
	a\cdot \omega &= \pi_L(a)\omega\label{prodl}\\
	\omega\cdot a &= \omega \pi_L(a) \label{prodr}
\end{align}
$a\in A_F$, and $\omega\in \Omega_D^1 A$, where
\begin{align}
	\omega&= \sum \begin{pmatrix}
		0 & \omega_1\otimes \omega_2^\ast\\
		\omega_3\otimes \omega_4^\ast&0\\
	\end{pmatrix},\label{1form}
\end{align}
with the sum taken over elements $\omega_i\in \mathbb{O}$. The product given in Eqs,~\eqref{prodl} and \eqref{prodr} are of split alternative form. We remind the reader that  in general $\pi_L(a)\pi_L(b)\neq \pi(ab)$, and so this action of $A_F$ on $\Omega^1_D A_F$ is not associative. 
The bimodule defined in  Eqs.~\eqref{prodl}, \eqref{prodr}, and \eqref{1form} is analogous to the split Jordan bimodule of  1-forms defined in~\cite{farnsworth2025npointexceptionaluniverse}, in which we considered finite dimensional geometries coordinatlized by exceptional Jordan algebras and with uncharged Dirac operators.

While the left and right  products given in \eqref{prodl} and \eqref{prodr} are compatible with derivations $\Delta_D$ established by the uncharged Dirac operator given in \eqref{unchargedD}, a key question is whether they are also compatible with the inner derivations of $A_F$ lifted to $H_F$. From Eqs.~\eqref{innerleftL2} and \eqref{innerleftR2} we have:
\begin{align}
	\delta_{ab}(c\cdot\omega) &= [\delta_{ab}',c\cdot \omega]=[\delta_{ab}',\pi_L(c) \omega]=[\delta_{ab}',\pi_L(c)] \omega+\pi_L(c)[\delta_{ab}', \omega]\nonumber\\
	&=\pi_L(\delta_{ab}c) \omega+\pi_L(c)\delta_{ab}(\omega) = \delta_{ab}(c)\cdot \omega+c\cdot\delta_{ab}(\omega),\\
	\delta_{ab}(\omega \cdot c) &= [\delta_{ab}',\omega\cdot c]=[\delta_{ab}', \omega\pi_L(c)]=[\delta_{ab}',\omega ]\pi_L(c)+\omega[\delta_{ab}',\pi_L(c)]\nonumber\\
	&=\delta_{ab}(\omega)\pi_L(c)+ \omega \pi_L(\delta_{ab}c)  =  \delta_{ab}(\omega)\cdot c+\omega\cdot \delta_{ab}(c).
\end{align}
$\delta_{ab}\in Der(A_F)$, $c\in A_F$, $\omega\in \Omega_D^1 A_F$. Our split alternative action of $A_F$ on $\Omega_D^1 A_F$ therefore fulfils the compatibility requirements given in Eqs.~\eqref{innerleftL} and \eqref{innerleftR}. In fact, as long as the  product between elements $a\in A_F$ and elements $\omega\in \Omega_D^1 A_F\subset  End(H_F)$ is constructed from sums of products of left and right acting operators $\pi_L(a),\pi_R(a)\in End(H_F)$, this will always be the case.


\subsection{The Construction of Charged Higgs Fields.}
\label{sec_charged}

In Sec.~\ref{sec_uncharged} we constructed a Dirac operator using only elements of the associative nucleus of the Octonions (i.e. scale multiples of the   identity element), and we saw that this Dirac operator was consistent with 1-forms defined as split alternative bimodules. We would like to construct finite dimensional, discrete, nonassociative spectral geometries corresponding to the internal spaces of gauge theories with charged Higgs fields. For this reason we return to the more general form of $D_F$ given in Eq.~\eqref{altDf}, along with the  Leibniz rule in Eq.~\eqref{leibrulealt}. Making use of the alternative identity given in Eq.~\eqref{eq_alt_2}, however, we re-express the Leibnize rule from Eq.~\eqref{leibrulealt} in a slightly  more general form:
\begin{align}
	\Delta_D(ab)	&= \Delta_D(a)\pi_L(b)+\pi_L(a)\Delta_D(b) +[D_F,(1-S)[\pi_L(a),\pi_R(b)] +S[\pi_R(a),\pi_L(b)]]\nonumber\\
	&= \Delta_D(a)\pi_L(b)+\pi_L(a)\Delta_D(b) +[D_F,(1-S)[\pi_L(a),J_F\pi_L(b)^*J_F] +S[J_F\pi_L(a)^*J_F,\pi_L(b)]],\label{leibrulealt2}
\end{align}
$a,b\in A_F$, and for some $S\in \mathbb{R}$. In the second equality we have made use of Eq.~\eqref{eq_right_action} to re-express the right action of algebra elements in terms of left acting elements and the involution operator $J_F$ on $H_F$. In order to reproduce a Leibniz rule of the form given in Eq.~\eqref{Deralt}, what we would like  is to re-express the RHS of Eq.~\eqref{leibrulealt2} purely in terms of elements  $\Delta_D(a),\Delta_D(b),L_a,$ and $L_b$. We note that the only  terms that obstruct us from doing so are those  of the form $[D_F,J_FL_a^*J_F]$ and $[D_F,J_FL_b^*J_F]$. To circumvent this obstruction,  notice that the Dirac operator $D_F$ can be broken  into two different kinds of terms $D_F = D_++D_-$:
\begin{align}
	D_+ &= M^{00}\begin{pmatrix}
		0&e_0\otimes e_0^\star\\
		e_0\otimes e_0^\star
	\end{pmatrix}+M^{ij}\begin{pmatrix}
		0&e_i\otimes e_j^\star\\
		e_j\otimes e_i^\star
	\end{pmatrix}\label{D+}\\
	D_- &= M^{0i}\begin{pmatrix}
		0&e_0\otimes e_i^\star\\
		e_i\otimes e_0^\star
	\end{pmatrix}+M^{i0}\begin{pmatrix}
		0&e_i\otimes e_0^\star\\
		e_0\otimes e_i^\star
	\end{pmatrix}\label{D-}
\end{align}
where  the $i,j$ range from $1$ to $7$, over the imaginary elements of $\mathbb{O}$. The components $D_\pm$ of $D_F$ then satisfy:
\begin{align}
	D_\pm J_F = \epsilon_\pm'J_FD_\pm,
\end{align}
for $\epsilon_\pm'= \pm 1$. If we restrict to a Dirac operator that satisfies the  constraint $D_FJ_F = \epsilon_F'J_F D_F$, for some choice $\epsilon_F' = \pm 1$, then this reduces $D_F$ to the form $D_\pm$ given in Eqs~\eqref{D+} and \eqref{D-}. For this restricted form of Dirac operator the Leibniz rule given in Eq.~\eqref{leibrulealt2} simplifies to:
\begin{align}
	\Delta_D(ab)	&= \Delta_D(a)\pi_L(b) +(1-S)[\Delta_D(a),J_F\pi_L(b)^*J_F] -\epsilon_F'S[J_F\Delta_D(a)^*J_F,\pi_L(b)],\nonumber\\
	&+\pi_L(a)\Delta_D(b) -\epsilon_F'(1-S)[\pi_L(a),J_F\Delta_D(b)^* J_F] +S[J_F\pi_L(a)^*J_F,\Delta_D(b)]. \label{edtendedaltleib1} 
\end{align}
Comparing equations Eq.~\eqref{edtendedaltleib1} and \eqref{Deralt}, we see that the Leibniz rule is then compatible with the following left and right bimodule actions:
\begin{align}
	\omega\cdot a	&= \omega\pi_L(a) +(1-S)[\omega,J_F\pi_L(a)^*J_F] -\epsilon_F'S[J_F\omega^*J_F,\pi_L(a)],\label{rightaction22}\\
	a\cdot \omega&=\pi_L(a)\omega -\epsilon_F'(1-S)[\pi_L(a),J_F\omega^* J_F] +S[J_F\pi_L(a)^*J_F,\omega],\label{leftaction22}
\end{align}
for $a\in A_F$, and $\omega\in \Omega_D^1 A_F$ of the general form given in Eq.~\eqref{1form}. If we further view $-\epsilon_F' J_F\omega^* J_F$ as the right action of the element $\omega\in \Omega_D^1 A_F$ on $H_F$, then we see that we have constructed an associative specialization rule for the representation of $A_F\oplus \Omega_D^1 A_F$ on $H_F$, analogous to that given in Eq.~\eqref{specialization}.

We have found an infinite family of bimodule  products parameterised by $S$, which are compatible with the derivations established by Hermitian Dirac operators satisfying $D_FJ_F = \epsilon_F'J_F D_F$, $\epsilon_F' = \pm 1$. Notice, however, that the coordinate algebra $A_F$ is an involutive algebra, and when defining the representation of $A_F$ on $H_F$ we extended the involution to $H_F$ by introducing the operator $J_F$ in Eq.~\eqref{InvolJ}. Similarly, we can   extended the involution to $\Omega_D^1A_F$ by introducing:
\begin{align}
	\omega^* = \omega^T,
\end{align}
for all $\omega\in \Omega_D^1 A_F$. This is the `obvious' involution to try, as one already has $\pi_L(a^*) = \pi_L(a)^T:= \pi_L(a)^*$, and $\pi_R(a^*) = \pi_R(a)^T:= \pi_R(a)^*$ for the representation of $A_F$ on $H_F$. Indeed, we can check that the transpose fulfils the role of an involution by seeing whether it can be used to relate the left and right actions:
\begin{align}
	\omega\star a  &= (a^*\star \omega^*)^*\nonumber\\
	&=	(\pi_L(a^*)\omega^T -\epsilon'(1-S)[\pi_L(a^*),J_F\omega J_F] +S[\pi_R(a^*),\omega^T])^T\nonumber\\
	&=	\omega\pi_L(a)+S[\omega,J_F\pi_L(a)^*J_F] -\epsilon'(1-S)[J_F\omega^* J_F,\pi_L(a)], 
\end{align}
where for the second equality we have made use of Eq.~\eqref{leftaction22}. Comparing with Eq.~\eqref{rightaction22},  we are forced to take $S = \frac{1}{2}$ if we wish for our representation to be involutive, with the involution given by the transpose. Not only does the involution of $A_F$ extend naturally to our bimodule under our  left and right bimodule actions, but we note as an aside that 
the identity element $\mathbb{I} = (e_0,e_0)\in A_F$ is also  compatible with these left and right products.

Our next task is to check whether the bimodule actions from Eqs.~\eqref{leftaction22} and \eqref{rightaction22} are  compatible with the continuous symmetries of $A_F$. In particular, we check that inner derivations $\delta_{ab}\in Der(A_F)$ lift appropriately to $\Omega_D^1 A_F$:
\begin{align}
	\delta_{ab}(c\cdot \omega)&=[\delta_{ab}',\pi_L(c)\omega -\epsilon'\frac{1}{2}[\pi_L(c),J_F\omega^* J_F] +\frac{1}{2}[J_F\pi_L(c)^*J_F,\omega] ]\nonumber\\
	&=\pi_L(\delta_{ab}c)\omega -\epsilon'\frac{1}{2}[\pi_L(\delta_{ab}c),J_F\omega^* J_F] +\frac{1}{2}[J_F\pi_L(\delta_{ab}c)^*J_F,\omega]\nonumber\\
	&+\pi_L(c)\delta_{ab}(\omega) -\epsilon'\frac{1}{2}[\pi_L(c),J_F\delta_{ab}(\omega)^* J_F] +\frac{1}{2}[J_F\pi_L(c)^*J_F,\delta_{ab}(\omega)] \nonumber\\
	&=\delta_{ab}(c)\cdot \omega + c\cdot \delta_{ab}(\omega)\\
	\delta_{ab}( \omega\cdot c)&=[\delta_{ab}', \omega\pi_L(c) +\frac{1}{2}[\omega,J_F\pi_L(c)^*J_F] -\epsilon'\frac{1}{2}[J_F\omega^*J_F,\pi_L(c)]]\nonumber\\
	&= \omega\pi_L(\delta_{ab}c) +\frac{1}{2}[\omega,J_F\pi_L(\delta_{ab}c)^*J_F] -\epsilon'\frac{1}{2}[J_F\omega^*J_F,\pi_L(\delta_{ab}c)]]\nonumber\\
	&+ \delta_{ab}(\omega)\pi_L(c) +\frac{1}{2}[\delta_{ab}(\omega),J_F\pi_L(c)^*J_F] -\epsilon'\frac{1}{2}[J_F\delta_{ab}(\omega)^*J_F,\pi_L(c)]]\nonumber\\
	&=\delta_{ab}( \omega)\cdot c+ \omega\cdot \delta_{ab}(c)
\end{align}
$c\in A_F$, $\omega\in \Omega_D^1 A_F$, and $\delta_{ab}\in Der(A_F)$. Here we have made use of the fact that derivations of the form given in Eq.~\eqref{inner_der_alt} commute with $J_F$, and with the involution such that $[\delta_{ab}',J]=0$ and $[\delta_{ab}',X^T] = [\delta_{ab}',X]^T$ for $X\in End(H_F)$. As expected, because the product  between elements $a\in A_F$ and elements $\omega\in \Omega_D^1 A_F\subset  End(H_F)$ is constructed from sums of products of left and right acting operators $\pi_L(a),\pi_R(a)\in End(H_F)$, we have automatic compatibility with the elements of $Der(A_F)$ lifted to $H_F$.

Finally, it is quick to check that $D_F$  of the form given in either Eqs. \eqref{D+} or \eqref{D-} will no longer commute with the lifted derivations of the geometry. As such, this internal space for either $\epsilon_F'=+1$ or $\epsilon_F'=-1$  will give rise to a gauge theory with charged Higgs fields. We construct this gauge theory in Appendix Section~\ref{Sec_quat_model}.


\section{Reconstituted Alternative Bimodules, External Derivations, And Automorphism Covariance}
\label{Sec_bimodules}

Motivated by the example geometry  in Sec \ref{sec_charged}, we now introduce a class of bimodules which we term \textit{reconstituted alternative bimodules}, defined over finite dimensional, semi-simple algernative algebras composed of octonionic factors. These bimodules are designed to admit external derivations that transform covarantly under the lifted automorphisms  of the algebra over which the bimodule is defined. Consider a finite-dimensional, unital, alternative algebra of the form 
$	A = \bigoplus_{i=1}^n  \mathbb{O}$, and define the corresponding bimodule $M$ as
\begin{align}
	M = \mathbb{O}\otimes (\bigoplus_{i,j=1}^n V^{ij})\otimes \mathbb{O} \label{eq-split_bi}
\end{align}
where each $V^{ij}$ is a real vector space with a canonical isomorphism $V^{ij} \cong V^{ji}$. For simplicity, we restrict to the case where each $V^{ij} \cong \mathbb{R}$, so that $M \cong A \otimes A$ as a vector space.
  We will call $M$ a reconstituted alternative bimodule over $A$  if it is equipped with the following left and right $\cdot$  products:
\begin{align}
	c_{(i)}\cdot (a\otimes^{(jk)} b)&=\hat{\delta}^j_i\frac{1}{2}\left[2(ca\otimes^{(jk)} b)+(ac\otimes^{(jk)}b)+\epsilon'(b\otimes^{(kj)} ac)\right]\nonumber\\
	&-\hat{\delta}^k_i\frac{1}{2}\left[(a\otimes^{(jk)} cb)+\epsilon'(cb\otimes^{(kj)} a)\right]\label{Eq_Charged_Left}\\
	(a\otimes^{(jk)} b)\cdot c_{(i)}&=\hat{\delta}^k_i\frac{1}{2}\left[2(a\otimes^{(jk)} bc)	+(a\otimes^{(jk)} cb)  + \epsilon'(cb\otimes^{(kj)} a)\right]\nonumber\\
	&-\hat{\delta}^j_i\frac{1}{2}\left[(ac\otimes^{(jk)} b)   + \epsilon'(b\otimes^{(kj)} ac)\right]\label{Eq_Charged_Right}
\end{align}
where $c_{(i)} = (0,...c,...,0)\in A$, with a single entry in the $i^{th}$ column, and  $\epsilon'=\pm 1$. The differential 1-forms discussed in Sec.~\ref{sec_charged} are a special case of this construction, where
\[
\Omega_D^1 A_F = \mathbb{O} \otimes \mathbb{R}^2 \otimes \mathbb{O}, \quad V^{12} = V^{21} = \mathbb{R}, \quad V^{11} = V^{22} = 0.
\]
A generic element $\omega \in \Omega_D^1 A$ in that context is structured as in Eq.~\eqref{1form}.

\subsection{Key Properties of Reconstituted Alternative Bimodules over Discrete Octonion Algebras}

Although the left and right bimodule actions  introduced in Eqs.~\eqref{Eq_Charged_Left} and ~\eqref{Eq_Charged_Right} are both noncommutative and nonassociative, they exhibit a number of desirable features. Chief among these is their compatibility with the identity element $\mathbb{I}=	(\sum_i e_{(i)0})\in A$, where each $e_{(i)0}$ denotes the unit of the $i$-th octonionic component. In particular:

\begin{align}
	(\sum_i e_0^{(i)})	\cdot (a\otimes^{(jk)} b)&= \frac{1}{2}\left[3(a\otimes^{(jk)} b)+\epsilon'(b\otimes^{(kj)} a)\right]-\frac{1}{2}\left[(a\otimes^{(jk)} b)+\epsilon'(b\otimes^{(kj)} a)\right]\nonumber\\
	&=a\otimes^{(jk)}b,\\
		 (a\otimes^{(jk)} b)\cdot (\sum_i e_0^{(i)})&= \frac{1}{2}\left[3(a\otimes^{(jk)} b)	  + \epsilon'(b\otimes^{(kj)} a)\right]-\frac{1}{2}\left[(a\otimes^{(jk)} b)   + \epsilon'(b\otimes^{(kj)} a)\right]\nonumber\\
	&=a\otimes^{(jk)}b.
\end{align}
In addition, the bimodule products support an involution:
\begin{align}
	(a\otimes^{(jk)}b)^* = b^*\otimes^{(kj)} a^*,\label{reconinvol}
\end{align}
which exchanges left and right actions via conjugation and transposition:
\begin{align}
	(c_{(i)}^*\cdot (a\otimes^{(jk)} b)^T)^T&=
	(c_{(i)}^*\cdot (b^*\otimes^{(kj)} a^*))^T\nonumber\\
	&=
	\hat{\delta}^k_i\frac{1}{2}\left[2(c^*b^*\otimes^{(kj)} a^*)+(b^*c^*\otimes^{(kj)}a^*)+\epsilon'(a^*\otimes^{(jk)} b^*c^*)\right]^T\nonumber\\
	&-\hat{\delta}^j_i\frac{1}{2}\left[(b^*\otimes^{(kj)} c^*a^*)+\epsilon'(c^*a^*\otimes^{(jk)} b^*)\right]^T\nonumber\\
	&=
\hat{\delta}^k_i\frac{1}{2}\left[2(a \otimes^{(jk)}bc)+(a\otimes^{(jk)}cb)+\epsilon'(cb\otimes^{(kj)} a)\right]-\hat{\delta}^j_i\frac{1}{2}\left[(ac\otimes^{(jk)} b)+\epsilon'(b\otimes^{(kj)}ac )\right]\nonumber\\
&= (a\otimes^{(jk)} b)\cdot c_{(i)}.
\end{align}
Crucially, as a bimodule, $M$ is also compatible with the continuous inner symmetries of $A$, where the lifted transformation of elements of $M$ is of the form:
\begin{align}
	\alpha'(a\otimes^{(jk)} b) = \alpha_j(a)\otimes^{(jk)} \alpha_{k}(b),\label{reconst_aut} 
\end{align}
where each $\alpha_j = \exp(\delta^j_{ab})$, with $\delta^j_{ab}$ an inner derivation of the alternative type (see Eq.~\eqref{Eq_Der_Alterative}) constructed from elements $a_{(j)}, b_{(j)} \in A$. In particular, one has:
\begin{align}
	\alpha(c_{(i)})\cdot \alpha'(a\otimes^{(jk)} b)&=\hat{\delta}^j_i\frac{1}{2}\left[2(\alpha(ca)\otimes^{(jk)} \alpha(b))+(\alpha(ac)\otimes^{(jk)}\alpha(b))+\epsilon'(\alpha(b)\otimes^{(kj)} \alpha(ac))\right]\nonumber\\
	&-\hat{\delta}^k_i\frac{1}{2}\left[(\alpha(a)\otimes^{(jk)} \alpha(cb))+\epsilon'(\alpha(cb)\otimes^{(kj)} \alpha(a))\right]\nonumber\\
	&=\alpha'(c_{(i)}\cdot(a\otimes^{(jk)} b) ),\\
	 \alpha'(a\otimes^{(jk)} b)\cdot \alpha (c_{(i)})	&=
	 \hat{\delta}^k_i\frac{1}{2}\left[2(\alpha (a) \otimes^{(jk)}\alpha (bc))+(\alpha (a)\otimes^{(jk)}\alpha (cb))+\epsilon'(\alpha (cb)\otimes^{(kj)} \alpha (a))\right]\nonumber\\
	 &-\hat{\delta}^j_i\frac{1}{2}\left[(\alpha (ac)\otimes^{(jk)} \alpha (b))+\epsilon'(\alpha (b)\otimes^{(kj)}\alpha (ac) )\right]\nonumber\\
	 &=\alpha'((a\otimes^{(jk)} b)\cdot c_{(i)}	).
\end{align}
Furthermore, the lifted automorphisms are compatible with the bimodule involution in the sense that they commute:
\begin{align}
	\alpha'((a\otimes^{(jk)}b)^*) &= \alpha'(b^*\otimes^{(kj)} a^*)\nonumber\\
	&= \alpha_k(b)^*\otimes^{(kj)} \alpha_j(a)^*\nonumber\\
	&= (\alpha_j(a)\otimes^{(jk)} \alpha_k(b))^*=(\alpha'(a\otimes^{(jk)}b))^*.
\end{align}
Finally, we note that the left and right bimodule actions given in Eqs.~\eqref{Eq_Charged_Left} and \eqref{Eq_Charged_Right} can be added together to form a symmetric action of the form:
\begin{align}
	c_{(i)}\cdot (a\otimes^{(jk)} b)
	+ (a\otimes^{(jk)} b)\cdot c_{(i)}
	&=\hat{\delta}^j_i(ca\otimes^{(jk)} b)+ \hat{\delta}^k_i(a\otimes^{(jk)} bc).
\end{align}
If the underlying product on $A = \bigoplus_i \mathbb{O}$ is replaced by the symmetrized octonionic product $a \circ b := \frac{1}{2}(ab + ba)$, then $A$ becomes a (special) Jordan algebra. In this case, symmetrizing the reconstituted alternative bimodule $B = A \oplus M$ with $M = A \otimes A$ results in a Jordan bimodule over a semi-simple special Jordan algebra.

\subsection{Derivation Compatible Bimodule Maps}

Our next goal is to demonstrate that one can construct automorphism-covariant external derivations on reconstituted alternative bimodules over semisimple Octonion algebras. Consider the Following Derivation $\Delta_\kappa:A\rightarrow M$, which is a slightly modified version of the derivation given in Eq.~\eqref{Eq_split_alt_der} for split alternative bimodules with vector spaces of the form given in Eq.~\eqref{eq-split_bi}:
\begin{align}
	\Delta_\kappa[e^{(a)i}] = \sum_c  \kappa_{(ac)}( e^i\otimes^{ac}  e^0 - e^0\otimes^{ca} e^i ),\label{Eq_uncharged_der_map}
\end{align}
where the coefficients are symmetric: $\kappa_{(ac)}=\kappa_{(ca)}$.
 In this expression, we have dropped the parentheses around superscripts over the tensor product, as they would become cumbersome in subsequent calculations. 
  Rather than deriving Eq.~\eqref{Eq_uncharged_der_map} from first principals, we instead verify that it is compatible with the  left and right bimodule actions defined in Eqs.\eqref{Eq_Charged_Left} and \eqref{Eq_Charged_Right}.  Specifically,  we check that  the Leibniz rule $\Delta_\kappa[e^{(a)i}e^{(b)j}] = \Delta_\kappa[e^{(a)i}]\cdot e^{(b)j}+e^{(a)i}\cdot \Delta_\kappa[e^{(b)j}]$ holds. On the LHS we have:
\begin{align}
LHS
&=\hat{\delta}^{ab}\sum_c  \kappa_{(ac)}( e^{i}e^{j}\otimes^{ac}  e^0 - e^0\otimes^{ca} e^{i}e^{j} )
\end{align}
On the RHS we have:
\begin{align}
	RHS&= 	\sum_c  \Big(\kappa_{(ac)} (e^i\otimes^{ac}  e^0 - e^0\otimes^{ca} e^i )\cdot e^{(b)j}+\kappa_{(bc)} e^{(a)i}\cdot ( e^j\otimes^{bc}  e^0 - \kappa_{(cb)}e^0\otimes^{cb} e^j) \Big)\nonumber\\
		&=	\sum_c  \Big(\frac{\kappa_{(ac)}}{2} (\hat{\delta}^c_b\left[2(e^i\otimes^{ac} e^j)	+(e^i\otimes^{ac}e^j)  + \epsilon'(e^j\otimes^{ca} e^i)\right]-\hat{\delta}^a_b\left[(e^ie^j\otimes^{ac} e^0)   + \epsilon'(e^0\otimes^{ca} e^ie^j)\right])\nonumber\\
		&- \frac{\kappa_{(ac)}}{2}(\hat{\delta}^a_b\left[2(e^0\otimes^{ca} e^ie^{j})	+(e^0\otimes^{ca} e^{j}e^i)  + \epsilon'(e^{j}e^i\otimes^{ac} e^0)\right]-\hat{\delta}^c_b\left[(e^{j}\otimes^{ca} e^i)   + \epsilon'(e^i\otimes^{ac} e^{j})\right])\nonumber\\
		&+\frac{\kappa_{(cb)}}{2} ( \hat{\delta}^b_a\left[2(e^ie^j\otimes^{bc} e^0)+(e^je^i\otimes^{bc}e^0)+\epsilon'(e^0\otimes^{cb} e^je^i)\right]-\hat{\delta}^c_a\left[(e^j\otimes^{bc} e^i)+\epsilon'(e^i\otimes^{cb} e^j)\right])\nonumber\\
		& - \frac{\kappa_{(cb)}}{2}(\hat{\delta}^c_a\left[2(e^i\otimes^{cb}  e^j)+(e^i\otimes^{cb} e^j)+\epsilon'( e^j\otimes^{bc} e^i)\right]-\hat{\delta}^b_a\left[(e^0\otimes^{cb} e^i e^j)+\epsilon'(e^i e^j\otimes^{bc}e^0)\right])\Big)\nonumber\\
						&=	\hat{\delta}^a_b\sum_c  \frac{\kappa_{(ac)}}{2}\Big( (\{e^i,e^j\}+\epsilon'[e^{i},e^j])\otimes^{ac} e^0 -e^0\otimes^{ca}( \{e^i,e^{j}\}	  + \epsilon'[e^i,e^j])\Big).
\end{align}
where we have repeatedly made use of the symmetric indices in the $\kappa_{ab}$. This confirms the validity of the Leibniz rule, but only when $\epsilon'=1$. As anticipated, the nonassociativity of the algebra does not enter this computation, since the derivation $\Delta_\kappa$ is constructed entirely from elements in the associative center of the octonions.
Moreover, we verify that these derivations are invariant under lifted automorphisms of the form given in Eq.~\eqref{reconst_aut},  exactly as occurs for split alternative bimodules:
\begin{align}
		\Delta_\kappa[\alpha(e^{(a)i})] &= \sum_c  \kappa_{(ac)}( \alpha(e^i)\otimes^{ac}  e^0 - e^0\otimes^{ca} \alpha(e^i) )\nonumber\\
		&=\alpha'\left( \sum_c  \kappa_{(ac)}( e^i\otimes^{ac}  e^0 - e^0\otimes^{ca} e^i )\right)= \alpha'(\Delta_\kappa[e^{(a)i}]).
\end{align}

\subsubsection{Automorphism Covariance When $\epsilon'= 1$}
 
 Let $M=A\otimes A$ be a reconstituted, alternative bimodule over a discret octonion algebra  $A=\bigoplus_1^n \mathbb{O}$. Our goal is to  construct  external  derivations $\Delta_\Phi: A\rightarrow M$ that transform  under the lifted automorphisms of $A$. As in the associative setting,  this can be achieved by composing an automorphism-invariant derivation with a bimodule map of the form
  \begin{align}
 	\Phi(\omega_1 \otimes \omega_2)= \sum(\omega_1 a\otimes b\omega_2), \label{altmaptran}
 \end{align}
where the sum is taken over elements $a,b\in A=\bigoplus_i^n \mathbb{O}$.
We focus on such maps because, just as in the associative setting, they transform naturally under the action of the lifted automorphisms of $A$. Specifically, for an automorphism $\alpha\in Aut(A)$ lifted to $M$ as in Eq.~\eqref{reconst_aut}, these maps transform as:
\begin{align}
	\Phi \mapsto \Phi' = (\alpha')^{-1} \circ \Phi \circ \alpha', \quad \text{so that} \quad \Phi'(\omega_1 \otimes \omega_2) = \sum (\omega_1 \alpha^{-1}(a) \otimes \alpha^{-1}(b) \omega_2).
\end{align}
Thus, provided $a,b\in A$ are not drawn from the associative nucleus of $A$, the resulting map $\Phi$ will transform nontrivially under $Aut(A)$, as desired. However, due to the nonassociativity of $A$, such maps do not in general satisfy the conditions for bimodule homomorphisms. For our purposes, this is not a problem. Instead, we only require that the composition $\Phi\circ\Delta_\kappa$ satisfy the weaker derivation compatibility condition
defined in Eq.\eqref{dericompat}, where  $\Delta_\kappa$ is the first order, automorphism-invariant derivation introduced in Eq.\eqref{Eq_uncharged_der_map}. Motivated by the example geometry of Sec.~\ref{sec_charged}, we now consider a specific class of transformations:
 \begin{align}
 	\Phi_{cd}(a\otimes^{ab} b) &= (S^{ab}_{cd})(ae^m\otimes^{ab} e^mb),\label{Eq-dercompatmap}
 \end{align}
where $S^{ab}_{cd} = \frac{1}{1+\hat{\delta}^{cd}}(\hat{\delta}^{ca}\hat{\delta}^{db}+\hat{\delta}^{cb}\hat{\delta}^{da})$. The operator $\Phi_{cd}\circ \Delta_\kappa$ is  a map from  $A$ into the space $\mathbb{O}\otimes (V^{cd} \oplus V^{dc} )\otimes \mathbb{O}\subset M$. In the above transformation, the choice of imaginary basis element $e^m$ on either side of the tensor product is arbitrary, except that the transformation must remain symmetric under the exchange of indices $c\leftrightarrow d$. This is because we can, without loss of generality, always redefine the basis for the algebra factors  $A_c$ and $A_d$ such that we can label the elements in the $\Phi_{cd}$  transformation `$e^m$' on both sides of the tensor product.
We will adopt such a choice to simplify upcoming calculations, as it allows us to avoid needing to track index symmetries within $\Phi_{cd}$.

To verify that $\Phi_{cd}\circ \Delta_\kappa$ satisfies the derivation compatibility condition in Eq.~\eqref{dericompat}, we compute both sides of the Leibniz rule given in Eq.~\eqref{dericompat}. On the left-hand side, we have:
\begin{align}
LHS&=	\Phi_{cd}(\Delta[e^{(a)i}e^{(b)j}])\nonumber\\
	&=\frac{\hat{\delta}^{ac}\hat{\delta}^{ab} \kappa_{(cd)}}{1+\hat{\delta}^{cd}} ( (e^{i}e^{j})e^m\otimes^{cd}  e^m - e^m\otimes^{dc} e^m(e^{i}e^{j}) )\nonumber\\
	&+\frac{\hat{\delta}^{ad}\hat{\delta}^{ab} \kappa_{(cd)}}{1+\hat{\delta}^{cd}} ( (e^{i}e^{j})e^m\otimes^{dc}  e^m - e^m\otimes^{cd} e^m(e^{i}e^{j}) )
\end{align}
$e^{(a)i},e^{(b)j}\in A$.  The corresponding right-hand side is, after a direct (but lengthy) computation:
 \begin{align}
	RHS	&=e^{(a)i}\cdot\Phi_{cd}(\Delta[e^{(b)j}])+\Phi_{cd}(\Delta[e^{(a)i}])\cdot e^{(b)j}\nonumber\\
&=	\frac{\hat{\delta}^{bc}  \hat{\delta}^{ca}\kappa_{(cd)}}{2(1+\hat{\delta}^{cd})}\Big(
	(2e^i (e^{j}e^m)+[e^{j},e^m,e^i]-[e^i,e^m,e^{j}])\otimes^{cd} e^m\nonumber\\
	&-e^m\otimes^{dc}( 2(e^me^{i})e^j)	- [e^j,e^m,e^{i}]+  [e^{i},e^m,e^j] ) 
	\Big)\nonumber\\
	&+\frac{\hat{\delta}^{bd}  \hat{\delta}^{da}\kappa_{(cd)}}{2(1+\hat{\delta}^{cd})}\Big(
	(2e^i(e^{j}e^m)+[e^{j},e^m,e^i]-[e^i,e^m,e^{j}])\otimes^{dc} e^m\nonumber\\
	&-e^m\otimes^{cd}( 2(e^me^{i})e^j	- [e^j,e^m,e^{i}]  + [e^{i},e^m,e^j])
	\Big)
	=LHS.
\end{align}
In the above expression the bracketed terms $[a,b,c]=(ab)c-a(bc)$ correspond to octonion associators, which are totally antisymmetric in their entries due to alternativity.  Comparing both sides, we see that the derivation compatibility condition is satisfied: $LHS=RHS$. Hence $\Phi_{cd}\circ\Delta$ defines a derivation. These derivation maps transform in the $(7,7)$ representation of the $G_2\times G_2$ automorphism subgroup corresponding to the $A_c\oplus A_d$ factors of the coordinate algebra.

\subsubsection{Automorphism Covariance When $\epsilon'= -1$}

Consider once again  the reconstituted, alternative bimodule $M =A\otimes A$, where  $A=\bigoplus_1^n \mathbb{O}$ is a semisimple octonion algebra. As we have seen, the external maps $\Delta_\kappa$ given in Eq.~\eqref{Eq_uncharged_der_map}  act as derivations only in the case $\epsilon' = 1$, for the bimodule actions given in Eqs.~\eqref{Eq_Charged_Left} and \eqref{Eq_Charged_Right}. This result  is consistent with the  example geometry discussed in Sec.~\ref{sec_charged}, where  the Dirac operator  in Eq.~\eqref{D-} could no longer be constructed purely from octonion identity elements  when  $\epsilon'=-1$. Nevertheless,  we  attempt to construct derivations of the form $\Phi_{cd}\circ \Delta_\kappa$, where $\Phi_{cd}$ is a module map and $\Delta_\kappa$ is as defined previously. Motivated by the form of the Dirac operator in Eq.~\eqref{D-}, we define transformations of the form:
\begin{align}
	\Phi_{cd}(a\otimes^{ab} b) &= (1-\hat{\delta}^{ab})\left( \hat{\delta}^{ca}\hat{\delta}^{db}(ae^m\otimes^{ab} b) -\hat{\delta}^{cb}\hat{\delta}^{da}(a\otimes^{ab} e^mb)\right),\label{Eq-dercompatmap-2}
\end{align}
where the sign is chosen so that $\Phi_{cd}$ maps anti-selfadjoint elements to selfadjoint elements as defined in terms of the involution  in Eq.~\eqref{reconinvol}.  The composition $\Phi_{cd}\circ \Delta$ defines  a map from  $A$ into the space $\mathbb{O}\otimes (V^{cd} \oplus V^{dc} )\otimes \mathbb{O}\subset M$. The choice of imaginary basis element $e^m$ is arbitrary. Without loss of generality we can always relabel the imaginary unit in the algebra factor  $A_c$ from which the  $\Phi_{cd}$  transformation is constructed. We now test  whether module maps of the form given in Eq.~\eqref{Eq-dercompatmap-2} are derivation compatible in the sense of Eq.~\eqref{dericompat}, when composed with $\Delta_\kappa$. On the LHS,  we compute:
 \begin{align}
 	LHS&=	\Phi_{cd}(\Delta_\kappa[e^{(a)i}e^{(b)j}])\nonumber\\
	&=\hat{\delta}^{ab}\hat{\delta}^{ca}  \kappa_{(ad)}\left( (e^{i}e^{j})e^m\otimes^{ad}  e^0 + e^0\otimes^{da} e^m(e^{i}e^{j})  \right)-\hat{\delta}^{ab} \hat{\delta}^{da} \kappa_{(ac)}\left(  e^{i}e^{j}\otimes^{ac}  e^m
 	+ e^m\otimes^{ca} e^{i}e^{j} \right)\nonumber\\
 	&-\hat{\delta}^{ab}   \kappa_{(aa)}\hat{\delta}^{ca}\hat{\delta}^{da}\left( (e^{i}e^{j})e^m\otimes^{aa}  e^0 + e^0\otimes^{aa} e^m(e^{i}e^{j}) -  e^{i}e^{j}\otimes^{aa}  e^m
 	- e^m\otimes^{aa} e^{i}e^{j} \right),
 \end{align}
 $e^{(a)i},e^{(b)j}\in A$. On the right-hand side, we compute:
 \begin{align}
 		RHS	&=e^{(a)i}\cdot\Phi_{cd}(\Delta[e^{(b)j}])+\Phi_{cd}(\Delta[e^{(a)i}])\cdot e^{(b)j}\nonumber\\
 		&=\frac{\kappa_{(ad)}
 			\hat{\delta}^{ca}}{2}\hat{\delta}^b_a\left[2e^i(e^je^m)+[e^j,e^m,e^i]-[e^i,e^m,e^j]+ (1+\epsilon')(e^j(e^me^i)-e^i(e^me^j))\right]\otimes^{ad} e^0\nonumber\\
 		&+
 		\frac{	\kappa_{(ad)}
 			\hat{\delta}^{ca}}{2}	 \hat{\delta}^a_be^0\otimes^{da}\left[2 (e^me^i)e^j	- [e^j,e^m,e^i]+[e^i,e^m,e^j]  
 		+(1+\epsilon') ((e^je^m)e^i- (e^ie^m)e^j)
 		\right]\nonumber\\
 		&-\frac{\kappa_{(ac)}
 			\hat{\delta}^{da}}{2}\hat{\delta}^b_a\left[(\{e^i,e^j\} +\epsilon'[e^j,e^i])\otimes^{ac} e^m+e^m\otimes^{ca}( \{e^i,e^j\}  +\epsilon' [e^j,e^i])\right]\nonumber\\
 		&+\frac{\kappa_{(bd)}
 			\hat{\delta}^{cb}}{2}\hat{\delta}^d_a(1+\epsilon')\left[(e^i\otimes^{ab}e^me^j)+(e^me^j\otimes^{ba} e^i)-(e^je^m\otimes^{ba} e^i)-(e^i\otimes^{ab} e^je^m)\right]\nonumber\\
 		&+
 		\frac{\kappa_{(ab)}
 			\hat{\delta}^{ca}}{2}	 \hat{\delta}^d_b(1+\epsilon')\left[(e^ie^m\otimes^{ab} e^j)  + (e^j\otimes^{ba} e^ie^m)-(e^j\otimes^{ba} e^me^i)   - (e^me^i\otimes^{ab} e^j)\right]\nonumber\\
 				 		 		&-\hat{\delta}^{cb}\hat{\delta}^{db}\kappa_{(bb)}e^{(a)i}\cdot\left( 
 		e^je^m\otimes^{bb}  e^0 + e^0\otimes^{bb} e^me^j -	 e^j\otimes^{bb}  e^m - e^m\otimes^{bb} e^j \right)\nonumber\\
 		&-
 		\hat{\delta}^{ca}\hat{\delta}^{da} \kappa_{(aa)}\left(
 		e^ie^m\otimes^{aa}  e^0 + e^0\otimes^{aa} e^me^i -	 e^i\otimes^{aa}  e^m - e^m\otimes^{aa} e^i \right)\cdot e^{(b)j}=LHS.
 	\end{align}
Using the alternativity of the octonions and the properties of the associators, the final equality $LHS=RHS$ holds precisely when  $\epsilon'=-1$. 
Thus, we confirm that the operator $\Phi_{cd}\circ\Delta$ defines a derivation in this case. These derivation maps transform in the $(7,1)$ representation of the $G_2\times G_2$ automorphism subgroup corresponding to the $A_c\oplus A_d$ factors of the coordinate algebra.

\section{Discussion}
\label{Sec_discussion}

We have now constructed examples of special Jordan~\cite{Farnsworth_2020,Farnsworth:2020ozj,Besnard_2022,Boyle:2020}, exceptional Jordan~\cite{farnsworth2025npointexceptionaluniverse,carotenuto2019}, and  alternative spectral geometries in the finite dimensional and discrete setting (see also~\cite{Wulky,Akrami_2004,Hassanzadeh_2015,carotenuto2019} for other works in nonassociative geometry). The  key insight that should be taken from these examples, and this paper is the following: The  challenge  of nonassociative spectral geometry boils down to solving the  equation:
\begin{align}
	[D,\pi_L(ab)]=[D,\pi_L(a)]\cdot \pi_L(b)+\pi_L(a)\cdot [D,\pi_L(b)],\label{keyey}
\end{align}
where $a,b\in A$, $D$ is an operator on $H$,  $\pi_L:A\rightarrow End(H)$ provides a nonassociative representation of $A$, and the data $T=(A,H,D)$ defines a spectral triple. Given a representation $\pi_L$, The task is to find a pair $(D,\cdot)$ such that  Eq.~\eqref{keyey} holds. This differs substantially from the associative setting where  the bimodule generated by elements $[D,\pi_L(a)]$  is assumed to be associative, and $D$ is completely unconstrained by Eq.~\eqref{keyey}.

Our attention to this point has focused primarily on  finite dimensional, semi-simple, nonassociative algebras that have representations which are known to satisfy associative specialization identities. That is, for  special Jordan, and alternative algebras one can find representations that satisfy respectively   $\pi_L(ab)=\frac{1}{2}\{\pi_L(a),\pi_L(b)\}$, and $\pi_L(ab) = \pi_L(a)\pi_L(b) + [\pi_L(a),\pi_R(b)]$. These specialization rules can be used to expand the LHS of Eq.~\eqref{keyey}, and thereby  to derive consistent pairs $(D,\cdot)$  in a relatively straightforward manner~\cite{Besnard_2022}.

More generally however, even if one does not have access to specialization identities, the challenge remains the same. All of the elements in Eq.~\eqref{keyey} are standard operators, and the form of the elements $\pi_L(a)\in End(H)$ is known for all $a\in A$. Furthermore, because one knows  that the Dirac operator must transform in a standard way under the lifted automorphisms of $A$, $D\rightarrow D'=\alpha'D(\alpha')^{-1}$, this means that Eq.~\eqref{keyey} must hold not just for $D$ but for any $D'$ in its orbit. This ensures that  Eq.~\eqref{keyey} is hugely constraining. This together with Hermiticity, and requiring $D$ to be `off-diagonal' in the discrete setting whittles down considerably the possibilities for $D$, ensuring that the challenge is far more tractable than it might initially seem. A key goal moving forwards will be to show that this procedure is indeed possible, with a coordinate algebra without associative specializations. The exceptional Jordan algebra provides an excellent candidate~\cite{Besnard_2022}, as do the Bison algebras~\cite{toro}, and Brown Algebras~\cite{gari1}.

Perhaps the most exciting aspect of this work from a mathematical perspective is that spectral geometry now offers a rich but constrained framework for exploring  nonassociative representation theory. The discrete internal spaces of gauge theory provide a physical sandbox for  deriving and exploring entirely novel bimodule structures, which as yet remains almost untouched.
 As we have demonstrated, when applied in the setting of  semisimple octonion algebras, one readily develops a notion of both split, and reconstituted alternative bimodules. It is not yet clear how far these notions might be generalized in the alternative setting.
 
 From a mathematical standpoint, there are also challenges to be solved:
\begin{enumerate}
	\item  The present discussion has been limited to first-order forms. Extending the framework to higher-order differential forms and exploring their properties and representations remains an open task.
	
	\item  The lessons above need also to be formalized into a coherent set of axioms for nonassociative spectral geometry, analogous to Connes' axioms in the associative case~\cite{connes96}, and as has been achieved in the special Jordan setting~\cite{Besnard_2022}.
	
\end{enumerate}

From a physical perspective, a compelling picture is taking shape. Nonassociative spectral geometry offers a powerful and highly constrained framework for modeling physical systems. In particular, it imposes strict conditions on allowable Dirac operators and bimodules, which in turn restrict the particle representations in gauge theories. Moreover, nonassociative algebras themselves offer intriguing prospects for model building: they possess exceptional symmetries~\cite{gari1,baez}, chiral representations~\cite{Fredy,toro}, and structures resembling particle generations and other features of the Standard Model~\cite{boylef4,DV:2016,tod2,todorov}. While we have initiated early efforts to identify the special Jordan internal geometries most closely related to the Standard Model~\cite{Boyle:2020,Besnard_2022}, the broader nonassociative setting remains largely unexplored. The results presented here provide a solid foundation for advancing that program.

\section*{Acknowledgements}

I would like to thank Felix Finster, Claudio Paganini, and Axel Kleinschmidt for their support during the
writing of this work. I would also like to thank Keegan Flood for useful discussions and
suggestions for improving this text.

\appendix 

\section{Derivation Bimodule Examples:}
\label{examplebimodules}

In this section we present two example derivation bimodules over associative algebras. As we will show, both of these examples fail the 
standard defining associative bimodule definitions given in Eqs.~\eqref{Eq_Assoc_Module_1}, \eqref{Eq_Assoc_Module_2}, and \eqref{Eq_Assoc_Module_3}. We also explain how these bimodules relate to the identities given in Eqs.~\eqref{weak3}, \eqref{weak1}, and \eqref{weak2}, which derive from the specific requirement that the derivation algebra $Der(B)$ for the bimodule $B=A\oplus M$ is of associative form as given in Eq.~\eqref{derassoclie}.

\subsection*{Example 1. Associative  algebras with `Lie' representation:} Consider the algebra $A=M_n(\mathbb{C})$ of $n\times n$ complex matricies represented on its own vector space via the following left and right actions:
\begin{align}
	\pi_L(a) &= \frac{1}{2}(L_a-R_a),\label{leftlieact}\\
	\pi_R(a) &= \frac{1}{2}(R_a-L_a)=-\pi_L(a)\label{rightlieact},
\end{align}  
$a\in A_F$. It is clear that $\pi_L$ and $\pi_R$ fail the usual properties of an associative representation. In particular:
\begin{align}
	\pi_L(ab) - \pi_L(a)\pi_L(b) &=  \frac{1}{4}(L_{ab}-2R_{ab}-R_aR_b+L_aR_b+R_aL_b)\neq 0\\
	\pi_R(ab) - \pi_R(b)\pi_R(a) &= \frac{1}{4}(R_{ab}-2L_{ab}-L_bL_a+L_bR_a+R_bL_a)\neq 0\\
	[\pi_L(a),\pi_R(b)]&=   \frac{1}{4}(R_{[a,b]}-L_{[a,b]} )\neq 0,
\end{align}
$a,b\in A$. It is easy to check, however, that the elements of $Der(A)$ lift naturally to $M$, and in particular one has lifted derivations of associative form $\delta'_a = \pi_L(a) - \pi_R(a)\in  End(M)$:
\begin{align}
	[\delta'_a,\pi_L(b)]&= [\pi_L(a)-\pi_R(a),\pi_L(b)]=\frac{1}{2}(L_{[a,b]}-R_{[a,b]}) = \pi_L(\delta_a b),\\
	[\delta'_a,\pi_R(b)]&= [\pi_L(a)-\pi_R(a),\pi_R(b)]=-\frac{1}{2}(L_{[a,b]}-R_{[a,b]}) = \pi_R(\delta_a b),
\end{align}
for all $a,b\in A$. Because the lifted derivations have associative form, it does indeed satisfy the conditions Eq.~\eqref{leftassoc}, \eqref{rightassoc}, and \eqref{weak1}, which derive from the existence of lifted Derivations of associative form. Unfortunately, one runs into trouble when trying to construct external derivations $\Delta:A\rightarrow M$, constructed from elements in $M$. For derivation $\Delta:A\rightarrow M$ to exist, they must satisfy the Leibniz rule $\Delta(ab) = \frac{1}{2}([\Delta(a),b]+[a,\Delta(b)])$, where the commutators result from the product between elements in $A$ and $M$ is given by Eqs.~\eqref{leftlieact} and \eqref{rightlieact}. The RHS of this expression is anti-symmetric under $a\leftrightarrow b$ exchange, while the LHS is not. For this reason there can be no external derivations, and it is quick to check that Eqs.~\eqref{weak3} and \eqref{weak2}, which derive from the existance of external derivations of the associative form given in Eq.~\eqref{derassoclie}, indeed fail. A similar story would have occured had we defined a Jordan, rather than anti-commutative action of $A=M_n(\mathbb{C})$ on its own vector space. Demanding the existence of external derivations of the associative form given in   Eq.~\eqref{derassoclie}, therefore excludes such representations, just as they are excluded by the usual defining properties of associative bimodules.

\subsection*{Example 2. Associative Algebras with Alternative Representation:} Consider the  quaternion algebra $\mathbb{H}$, represented on a copy of its own vector space  via the following left and right actions:
\begin{align}
	\pi_L(a) &= R_a,\label{quatl}\\
	\pi_R(a) &= R_{a^*},\label{quatr}
\end{align}
$a\in \mathbb{H}$, and where the $\ast$ is the involution on $\mathbb{H}$. It is easy to check that this representation fails the usual properties of an associative bimodule. In particular:
\begin{align}
	\pi_L(ab) - \pi_L(a)\pi_L(b) &= R_{[a,b]}\neq0,\\
	\pi_R(ba) -\pi_R(a)\pi_R(b) &= R_{[a^*,b^*]}\neq0,\\
	[\pi_R(a),\pi_L(b)]&= R_{[b,a^*]}\neq0,\\
	[\pi_L(a),\pi_R(b)]&=  R_{[b^*,a]}\neq0,
\end{align}
$a,b\in\mathbb{H}$.  Using these expressions, however, it is  easy to show that the  left and right actions given in Eqs.~\eqref{quatl} and \eqref{quatr} automatically satisfy our weaker derived conditions in Eqs.~\eqref{weak1} and \eqref{weak2}:
\begin{align}
	\pi_L(ab) - \pi_L(a)\pi_L(b) - \pi_R(ba) +\pi_R(a)\pi_R(b) &= R_{[a,b]}- R_{[a^*,b^*]}=0\\
	[\pi_R(a),\pi_L(b)]-[\pi_L(a),\pi_R(b)] &= R_{[b,a^*]}- R_{[b^*,a]} = 0
\end{align}
for all $a,b\in\mathbb{H}$. The elements of $Der(A)$ also lift naturally to $M$, with $\delta'_a = -\frac{1}{2}(\pi_L(a) - \pi_R(a))\in Der(B)\subset End(M)$:
\begin{align}
	[\delta'_a,\pi_L(b)]&= -\frac{1}{2}[R_a-R_{a^*}, R_b]=-[R_a, R_b] = -R_{[b,a]} = \pi_L(\delta_a b)\\
	[\delta'_a,\pi_R(b)]&= -\frac{1}{2}[R_a-R_{a^*}, R_{b^*}]=-[R_a, R_{b^*}] = -R_{[b^*,a]} =-R_{[a^*,b]^*} = \pi_R(\delta_a b)
\end{align}
for $a,b\in A$. while the lifting conditions given in Eqs.~\eqref{lift1} and \eqref{lift2} are satisfied, they are only satisfied for  derivation operators $\delta_a'$, which are a scale multiple of the associative form given in Eq.~\eqref{derassoclie}. Similarly, external derivations $\Delta:A\rightarrow M$ exist, but they are not of the associative form given in Eq.~\eqref{derassoclie} at all. They are instead of the alternative form given in Eq.~\eqref{Eq_Der_Alterative}:
\begin{align}
	\Delta_{a,h} = [L_a,L_h]+[L_a,R_h]+[R_a,R_h],
\end{align}
$a\in A$, $h\in M$. In fact, all elements of $Der(B)$ including those in the subalgebra $Der(A)$ can be written in alternative form. Consequently, it is quick to check that the associativity condition given in Eq.~\eqref{weak3}, which derives from the associative derivation form fails,  as do Eqs.~\eqref{leftassoc} and \eqref{rightassoc}. We see that by allowing for a slightly more general alternativive form for the derivations $Der(B)$, we gain access to new kinds of bimodules, which preserve the symmetries of the algebra and  have interesting external derivations $\Delta:A\rightarrow M$, but which would usually be excluded by the standard definition of associative bimodules.

\section{The Octonion Algebra}

The octonions are discussed at length elsewhere~\cite{Furey_2018,Barnes:2016cjm,baez}. Here we provide the bare minimum required to understand this paper for someone who is completely unfamiliar with their construction. The octonion algebra $\mathbb{O}$ is a real 8-dimensional algebra that generalize the quaternions, just as the quaternions generalize the complex numbers. 
 The octonion basis includes a unit $e_0$, and 7 `imaginary' elements denoted $e_i$, $i=1,...,7$. The product between the imaginary elements is neatly captured by the Fano plane:
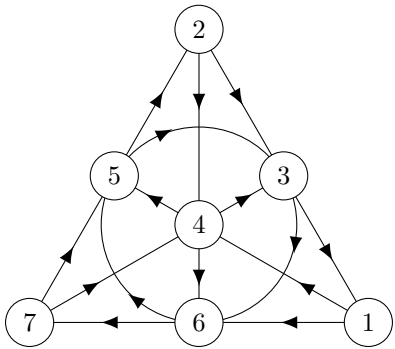
\begin{center}
	
	\begin{tikzpicture}[scale=1.3,cap=round,>=latex]
		\draw[
		decoration={markings, mark = between positions .26 and 0.95 step 0.53 with {\arrow[very thick]{>}}},
		postaction={decorate}
		] (210:2)--(90:2)  ;
		\draw[
		decoration={markings, mark = between positions .26 and 0.95 step 0.53 with {\arrow[very thick]{>}}},
		postaction={decorate}
		] (90:2)  --(330:2);
		\draw[
		decoration={markings, mark = between positions .26 and 0.95 step 0.53 with {\arrow[very thick]{>}}},
		postaction={decorate}
		] (330:2)--(210:2);
		
		\draw[
		decoration={markings, mark = between positions .18 and 0.95 step 0.6 with {\arrow[very thick]{<}}},
		postaction={decorate}
		]  (30:1)  -- (210:2);
		\draw[
		decoration={markings, mark = between positions .18 and 0.95 step 0.6 with {\arrow[very thick]{<}}},
		postaction={decorate}
		]  (150:1) -- (330:2);
		\draw[
		decoration={markings, mark = between positions .18 and 0.95 step 0.6 with {\arrow[very thick]{<}}},
		postaction={decorate}
		]  (270:1) -- (90:2);
		
		\draw[
		decoration={markings, mark = between positions .32 and 0.99 step 0.33 with {\arrow[very thick]{<}}},
		postaction={decorate}
		]
		(0,0) circle (1);
		\filldraw[fill=white, draw=black] 
		(0:0)   circle(7pt)
		(30:1)  circle(7pt)
		(90:2)  circle(7pt)
		(150:1) circle(7pt)
		(210:2) circle(7pt)
		(270:1) circle(7pt)
		(330:2) circle(7pt);
		\node[circle] at (0:0) {4};
		\node[circle] at (30:1) {3};
		\node[circle] at (90:2) {2};
		\node[circle] at (150:1) {5};
		\node[circle] at (210:2) {7};
		\node[circle] at (270:1) {6};
		\node[circle] at (330:2) {1};
	\end{tikzpicture}
\end{center}
The  product is read off from the above Fano plane diagram by following the arrows in each closed loop of three elements, with the product picking up a minus sign when going against the direction of the arrows. For example, $e_2e_3 = e_1$, $e_7e_6=-e_1$, $e_3e_5=-e_6$. 

The octonions are nonassociative, but they satisfy the properties of an alternative algebra, which means that the associator of elements $[a,b,c]=(ab)c-a(bc)$ for $a,b,c\in\mathbb{O}$ changes sign under the exchange of any two elements. The octonions are also involutive, which means that they are equipped with an involution $\ast:A\rightarrow A$ given by $e_0^* = e_0$ and $e_i^* = -e_i$ for all $i=1,...,7$. The involution satisfies the usual properties $(ab)^* = b^*a^*$ for $a,b\in \mathbb{O}$.

The octonions are also an example of what is known as a normed algebra, which means that they are equipped with a norm \(\|x\| = \sqrt{x x^*}\), where norm satisfies \(\|xy\| = \|x\| \cdot \|y\|\) for all \(x, y\).  In particular, this implies that the octonions form a division algebra, since \(\|x\| \neq 0\) for \(x \neq 0\) guarantees the existence of a two-sided inverse given by \(x^{-1} = x^* / \|x\|^2\).

\section{The Octonions As A Hilbert Space}
\label{Sec_hilmb}

The octonion algebra has a natural left representation on itself, which allows us to view the 8-dimensional real vector space of octonions as a left alternative module over itself.    We denote this space by $H = \mathbb{R}^8$.  That is, we identify the representation map $\pi(a) = L_a$, where $L_ah = ah$ for $a\in \mathbb{O}$, $h\in H$, such that  $\pi(a)\in End(H)$. 

A trace form can be defined on the octonions  by:
\begin{align}
	Tr[a] = a_0
\end{align} 
for $a = \sum_{m=0}^7a_Ie^I$. This trace form is symmetric: $Tr[ab]=Tr[ba]$, and is proportional to the operator trace on the left action of the algebra on itself $Tr[a] =\frac{1}{8} Tr[\pi(a)]$. The trace form can be used to define a real inner product on the vector space of octonions:
\begin{align}
	\langle a|b\rangle &:=Tr[a^*b]\\
	&=\sum_{I=0}^7a_Ib_I 
\end{align}
for $a,b\in\mathbb{O}$. This inner product reproduces the standard Euclidean inner product on $\mathbb{R}^8$.

In this paper we  construct spectral triples of the form $T = (A,H,D)$, where $A = \oplus_a^n A_a$, with each $A_a$  a copy of the octonion algebra $\mathbb{O}$, and where the Hilbert space is given by $H = \mathbb{O}\otimes \mathbb{R}^n$, with a coordinatewise action. The inner product on $H$ is defined by:
\begin{align}
	\langle a| b\rangle = \sum_a^n Tr[a_a^*b_a]
\end{align}
for $a=(a_1,...,a_n),b=(b_1,...,b_n)\in A$. The Hilbert space $H$ acts as an alternative left module over $A$ with the following  action:
\begin{align}
	\pi_L(a)h = \sum_a^n L_{a_a}h_a
\end{align}
$a=(a_1,...,a_n)\in A$, $h=(h_1,...,h_n)$. One can similarly define a `right' action:
\begin{align}
	\pi_R(a)h = \sum_a^n R_{a_a}h_a.
\end{align}

Because the  left and right action on the octonions  are given by linear operators, they can be expressed in the usual way:
\begin{align}
	L_a &= (L_a)^{mn}e_m\otimes e_n^\star,\\
	R_a &= (R_a)^{mn}e_m\otimes e_n^\star,
\end{align}
where the superscript $\star$ denotes the dual with respect to the inner product on $\mathbb{R}^8$,  such that $e_m^\star(e_n) = \hat{\delta}^n_m$, where $\hat{\delta}^{ab}$ denotes the Kronecker delta. We then have:
\begin{align}
	b^\star L_a &=  (a^* b)^\star\\
	b^\star R_a &=  ( ba^*)^\star
\end{align}
for $a,b\in\mathbb{O}$. Care should be taken not to confuse the superscripts $\ast$ and $\star$, which denote the involution and the dual, respectively.

\section{Identifying $\Omega_\Delta^1 A$ as a Split Alternative Bimodule over $A$}
\label{sec_identify}

Consider a finite dimensional, real, alternative algebra $\bigoplus_i^n A_i$, where each factor $A_i$ is a copy of the octonions $\mathbb{O}$. Let $M=A\otimes A$ is a split alternative bimodule over $A$. Define the sub-bimodule $\Omega_\Delta^1 A$ as the one  generated by elements of the form
\begin{align}
	\Delta_{\mathbb{I}\otimes \mathbb{I}}[a] &=a\otimes \mathbb{I}- \mathbb{I}\otimes a, & \forall a&\in A .
\end{align}
Our goal is to determine the structure of $\Omega_\Delta^1 A$. A well known result about the octonion algebra is that its left  multiplication algebra is isomorphic to $End(\mathbb{R}^8)$. In particular, any operator in $End(\mathbb{O})$ can be expressed as a linear combination of nested left multiplications~\cite{Furey_2018}:
\begin{align}
	M = c_0L_{e_0} + c_iL_{e_i}+... c_{123456}L_{e_1}L_{e_2}L_{e_2}L_{e_3}L_{e_4}L_{e_5}L_{e_6}.
\end{align}
where the $L_{e^i}$ are left multiplication operators by the standard basis elements of $\mathbb{O}$, and $c_I\in \mathbb{R}$. This representation also holds for any inner derivation $\delta_{ab}\in Der(\mathbb{O})$,  which can therefore be built from left multiplications alone. We define the following lift of derivations to $M$, constructed purely in terms of  nested left multiplication of elements in $A$ on elements in $M$:
\begin{align}
	\delta_{cd}'(\omega_1\otimes\omega_2):=\delta_{cd}(\omega_1)\otimes \omega_2,
\end{align}
$\delta_{cd} = (\delta_{cd}^1,...,\delta_{cd}^n)\in Der(A)$, $\omega_1\otimes\omega_2\in M$. Since inner derivations annihilate the identity, we have:
\begin{align}
	\delta_{cd}'(\omega_1 \otimes \omega_2) := \delta_{cd}(\omega_1) \otimes \omega_2, \qquad \omega_1 \otimes \omega_2 \in M.
\end{align}
Since inner derivations annihilate the identity, we have:
\begin{align}
	\delta_{cd}'(\Delta_{\mathbb{I} \otimes \mathbb{I}}[a]) = \delta_{cd}(a) \otimes \mathbb{I}, \qquad \forall a \in A.
\end{align}
Because the octonions form a division algebra, any nonzero element is invertible. For $a\in A$ such that each component $a_i \in A_i$ 
has nonzero imaginary part, the element $\delta_{cd}(a)$ is invertible. Hence, we can recover the $\mathbb{I}\otimes\mathbb{I}\in M$:
\begin{align}
	(\delta_{cd}(a))^{-1}\cdot \delta_{cd}'(	\Delta_{\mathbb{I}\otimes \mathbb{I}}[a])=\mathbb{I}\otimes\mathbb{I},
\end{align}
Finally, since $M$  is a split bimodule, we can generate any element $b\otimes d\in M$ by left and right multiplying  $\mathbb{I}\otimes\mathbb{I}\in M$ by any desired elements:
\begin{align}
	b \otimes d = b \cdot \left( (\delta_{cd}(a))^{-1} \cdot \delta_{cd}'(\Delta_{\mathbb{I} \otimes \mathbb{I}}[a]) \right) \cdot d.
\end{align}
Thus, elements of the form $\Delta_{\mathbb{I}\otimes \mathbb{I}}[a]$ generate all of $A\otimes A$ as an $A$-bimodule. We conclude that
\begin{align}
	\Omega_\Delta^1 A = A \otimes A.
\end{align}

\section{The Construction Of Gauge Theories From Spectral Geometry.}
\label{Sec_quat_model}

In Sec.~\ref{sec_example_geometry} we constructed the finite dimensional, internal geometry  $T_F= (A_F,H_F,D_F)$ corresponding to two  distinct $G_2\times G_2$ gauge theories with  different Higgs content. In this section we explain how to   construct the external commutative   geometry $T_c$ and how to form the full product geometry $T=T_c\otimes T_F$, which encodes the complete gauge theory. The external geometry $T_c$ is defined  by the commutative algebra of smooth real functions $A_c=C^\infty(M,\mathbb{R})$ over a Riemannian spin manifold $M$~\cite{Cacic}. This algebra is  represented on the Hilbert space of square integrable  spinors $H_c = L^2(M,S)$, and the Dirac operator is  the standard curved space Dirac operator
\begin{align}
	D_c = e^\mu_a \gamma^a \nabla^S_\mu,
\end{align}
where $e^\mu_a$ is the vierbein, $\gamma^a$ are the gamma matrices, and $\nabla^S_\mu$ is the spin connection. The gamma matrices are given in the Weyl basis by:
\begin{align}
	\gamma^0 &= \begin{pmatrix}
		0&\mathbb{I}\\
		\mathbb{I}&0
	\end{pmatrix}, & 	\gamma^1 &= \begin{pmatrix}
		0&i\sigma^1\\
		-i\sigma^1&0
	\end{pmatrix}, & 	\gamma^2 &= \begin{pmatrix}
	0&i\sigma^2\\
	-i\sigma^2&0
\end{pmatrix}, &	i\gamma^3 &= \begin{pmatrix}
		0&\sigma^3\\
		-i\sigma^3&0
	\end{pmatrix},
\end{align}
From these we  construct real structure and grading operators for the continuous space
\begin{align}
	J_c &= \gamma^0\gamma^2\circ cc, & \gamma_5 &= \gamma^0\gamma^1\gamma^2\gamma^3,
\end{align}
where \text{cc} denotes complex conjugation. These operators satisfy the following relations characteristic of KO-dimension 4~\cite{ConnesJ}:
\begin{align}
	J_c^2 &= \epsilon_c\mathbb{I}\\
	J_cD_c &=\epsilon_c'D_cJ_c\\
J_c\gamma_c &=\epsilon_c''\gamma_cJ_c	 	
\end{align}
with $\{\epsilon_c,\epsilon_c',\epsilon_c''\}=\{-1,-1,1\}$.
We now form the product geometry $T = (A, H, D)$ via:
\begin{align}
	A &= C^\infty(M,A_F), &H& = H_c\otimes H_F, & D = D_c\otimes \mathbb{I} + \gamma_5\otimes D_F,
\end{align}
with the tensor product  taken over $\mathbb{R}$.  Real structure and  grading operators are also defined on the total Hilbert space:
\begin{align}
	J &= \frac{1}{2}(1-\epsilon_F')J_c\otimes J_F+\frac{1}{2}(1+\epsilon_F')\gamma_cJ_c\otimes J_F, & \gamma = \gamma_5\otimes \begin{pmatrix}
		L_{e_0}&0\\
		0& -L_{e_0}
	\end{pmatrix}.
\end{align}
These new operators also satisfy real structure conditions $J^2 = \epsilon \mathbb{I}$, $JD =\epsilon'DJ$, and $J\gamma =\epsilon''\gamma J$, with $\{\epsilon_c,\epsilon_c',\epsilon_c''\}=\{-1,\epsilon_F,1\}$, corresponding once again to  KO-dimension 4 (the internal space $T_F$ is KO-dimension 0). 

To ensure gauge invariance, the Dirac operator is ``fluctuated'' by inner automorphisms:
\begin{align}
	D \rightarrow D + F,
\end{align}
where the fluctuation term $F$ captures the gauge and Higgs fields. We propose the following form for $F$:
 \begin{align}
 	F = \sum_{a,b}[[D,\pi_L(a)],\pi_L(b)+\frac{1}{2}J\pi_L(b)^*J]- \epsilon'[J[D,\pi_L(a)]^*J,J\pi_L(b)^*J+ \frac{1}{2}\pi_L(b)] \label{fluc} 
 \end{align}
where $a, b \in A$ are purely imaginary elements. This form ensures that $F$ is Hermitian and satisfies the compatibility condition $FJ = \epsilon' JF$, mirroring the corresponding relation for $D$. This ensures that the fluctuations of the internal part of the Dirac operator (i.e. the Higgs fields) act as Derivations, while the relative coefficients `$\frac{1}{2}$' that appear in the expression are chosen to ensure that the gauge connection terms act as  derivations.  Substituting the explicit form for the total Dirac operator into Eq.~\eqref{fluc} yields the following:
\begin{align}
	F &= \sum_{a,b}ie^\mu_a\gamma^a \partial_\mu(a^{(c)i})b^{(d)j} \otimes \delta'_{e_{(c)i}e_{(d)j}} \nonumber\\
	&+\gamma_5 a^{(c)i}b^{(d)j}\otimes\left([[D_F,L_{e_{(c)i}}],L_{e_{(c)i}}+\frac{1}{2}R_{e_{(c)i}}]+[[D_F,R_{e_{(c)i}}],R_{e_{(c)i}}+\frac{1}{2}L_{e_{(c)i}}]\right),
\end{align}
which can be expressed more compactly as:
\begin{align}
	F = \sum ie^\mu_a \gamma^a F(x)_\mu^X\otimes \delta_X +\gamma_5\Phi_{IJ}(x)\otimes 
	\begin{pmatrix}
		0 & e_I\otimes e_J^\ast\\
		e_J\otimes e_I^\ast&0\\
	\end{pmatrix},
\end{align}
where the $F(x)_\mu^i$ are real coefficients, and the $\delta_X$ form a basis for the inner derivations given in Eq.~\eqref{inner_der_alt}. The cofficients $\Phi_{IJ}$ are constrained such that $FJ=\epsilon'JF$. The first term in $F$ corresponds to gauge fields, which generate the $SU_2\times SU_2$ gauge symmetry, while the second term corresponds to Higgs fields $\Phi$, with the full covariant Dirac operator  given by $D+F$. With an appropriate action, constructed as a functional of $D+F$, one then reproduces the dynamics of the corresponding  gauge theory~\cite{Chamseddine_1997}.

\end{document}